\documentclass[12pt,preprint]{aastex}

\newcommand{\feII}{\ion{Fe}{2} $\lambda$5169}
\newcommand{\fe}{\ion{Fe}{2}}
\newcommand{\VmI}{\emph{V}$-$\emph{I}}
\newcommand{\vel}{$v_{\mathrm{Fe\,II}}$}

\usepackage{graphicx}
\usepackage{color}
\usepackage{amsmath,amssymb}
\usepackage{subfig}

\begin{document}

\title{Type II-P Supernovae from the SDSS-II Supernova Survey and the Standardized Candle Method\altaffilmark{1}}

\shorttitle{SDSS-II SN Survey: SNe\,II-P Standardization}
\shortauthors{D'Andrea et al.}

\author{Chris B. D'Andrea\altaffilmark{2},
Masao Sako\altaffilmark{2}, 
Benjamin Dilday\altaffilmark{3},
Joshua A. Frieman\altaffilmark{4,5,6}, 
Jon Holtzman\altaffilmark{7}, 
Richard Kessler\altaffilmark{4,5}, 
Kohki Konishi\altaffilmark{8}, 
D. P. Schneider\altaffilmark{9}, 
Jesper Sollerman\altaffilmark{10,11}, 
J. Craig Wheeler\altaffilmark{12}, 
Naoki Yasuda\altaffilmark{8}, 
David Cinabro\altaffilmark{13}, 
Saurabh Jha\altaffilmark{3},
Robert C. Nichol\altaffilmark{14}, 
Hubert Lampeitl\altaffilmark{14},
Mathew Smith\altaffilmark{14,15}, 
David W. Atlee\altaffilmark{16},
Bruce Bassett\altaffilmark{15,17},
Francisco J. Castander\altaffilmark{18},
Ariel Goobar\altaffilmark{19,20},
Ramon Miquel\altaffilmark{21,22}
Jakob Nordin\altaffilmark{19,20},
Linda \"Ostman\altaffilmark{19,20},
Jose Luis Prieto\altaffilmark{23,24},
Robert Quimby\altaffilmark{25}, 
Adam G. Riess\altaffilmark{26,27}, and
Maximilian Stritzinger\altaffilmark{10,28}}

\email{cdandrea@physics.upenn.edu}
\altaffiltext{1}{Based in part on data collected at Subaru Telescope, which is operated by the National Astronomical Observatory of Japan.}
\altaffiltext{2}{Department of Physics and Astronomy, University of Pennsylvania, 209 South 33rd Street, Philadelphia, PA  19104}
\altaffiltext{3}{Department of Physics and Astronomy, Rutgers University, 136 Frelinghuysen Road, Piscataway, NJ  08854}
\altaffiltext{4}{Kavli Institute for Cosmological Physics, The University of Chicago, 5640 South Ellise Avenue, Chicago IL  60637}
\altaffiltext{5}{Department of Astronomy and Astrophysics, The University of Chicago, 5640 South Ellise Avenue, Chicago IL  60637}
\altaffiltext{6}{Center for Particle Astrophysics, Fermi National Accelerator Laboratory, P.O. Box 500, Batavia, IL  60510}
\altaffiltext{7}{Department of Astronomy, MSC 4500, New Mexico State University, P.O. Box 30001, Las Cruces, NM  88003}
\altaffiltext{8}{Institute for Cosmic Ray Research, University of Tokyo, 5-1-5, Kashiwanoha, Kashiwa, Chiba, 277-8582, Japan}
\altaffiltext{9}{Department of Astronomy and Astrophysics, 525 Davey Laboratory, Pennsylvania State University, University Park, PA  16802}
\altaffiltext{10}{Dark Cosmology Centre, Niels Bohr Institute, University of Copenhagen, Juliane Maries Vej 30, DK-2100 Copenhagen \O, Denmark}
\altaffiltext{11}{The Oskar Klein Centre, Department of Astronomy, AlbaNova, Stockholm University, SE-106 91 Stockholm, Sweden}
\altaffiltext{12}{Department of Astronomy, McDonald Observatory, University of Texas, Austin, TX  78712}
\altaffiltext{13}{Department of Physics and Astronomy, Wayne State University, Detroit, MI  48202}
\altaffiltext{14}{Institute of Cosmology and Gravitation, Dennis Sciama Building, Burnaby Road, University of Portsmouth, Portsmouth PO1 3FX, UK}
\altaffiltext{15}{Department of Mathematics and Applied Mathematics, University of Cape Town, Rondebosch 7701, South Africa}
\altaffiltext{16}{Department of Astronomy, The Ohio State University, 140 West 18th Avenue, Columbus, OH  43210}
\altaffiltext{17}{South African Astronomical Observatory, P.O. Box 9, Observatory 7935, South Africa}
\altaffiltext{18}{Institut de Ci\`encies de l'Espai, (IEEC - CSIC), Campus UAB, 08193 Bellaterra, Barcelona, Spain}
\altaffiltext{19}{Department of Physics, Stockholm University, Albanova University Center, SE-106 91 Stockholm, Sweden}
\altaffiltext{20}{The Oskar Klein Centre for Cosmoparticle Physics, Department of Physics, AlbaNova, Stockholm University, SE-106 91 Stockholm, Sweden}
\altaffiltext{21}{Instituci\'o Catalana de Recerca i Estudis Avan\c{c}ats, Barcelona, Spain}
\altaffiltext{22}{Institut de F\'{\i}sica d'Altes Energies, E-08193 Bellaterra, Barcelona, Spain}
\altaffiltext{23}{Carnegie Observatories, 813 Santa Barbara St., Pasadena, CA  91101}
\altaffiltext{24}{Hubble and Carnegie-Princeton Fellow}
\altaffiltext{25}{Department of Astronomy, California Institute of Technology, Pasadena, CA  91125}
\altaffiltext{26}{Space Telescope Science Institute, 3700 San Martin Drive, Baltimore, MD  21218}
\altaffiltext{27}{Department of Physics and Astronomy, Johns Hopkins University, 3400 North Charles Street, Baltimore, MD  21218}
\altaffiltext{28}{Las Campanas Observatory, Carnegie Observatories, Casilla 601, La Serena, Chile}

\begin{abstract}
We apply the Standardized Candle Method (SCM) for Type II Plateau supernovae (SNe II-P), which relates the velocity of the ejecta of a SN to its luminosity during the plateau, to 15 SNe II-P discovered over the three season run of the Sloan Digital Sky Survey - II Supernova Survey.   The redshifts of these SNe - $0.027 < z < 0.144$ - cover a range hitherto sparsely sampled in the literature; in particular, our SNe II-P sample contains nearly as many SNe in the Hubble flow ($z > 0.01$) as all of the current literature on the SCM combined.   We find that the SDSS SNe have a very small intrinsic \emph{I}-band dispersion ($0.22$ mag), which can be attributed to selection effects.  When the SCM is applied to the combined SDSS-plus-literature set of SNe II-P, the dispersion increases to $0.29$ mag, larger than the scatter for either set of SNe separately.  We show that the standardization cannot be further improved by eliminating SNe with positive plateau decline rates, as proposed in \citet{P09}.  We thoroughly examine all potential systematic effects and conclude that for the SCM to be useful for cosmology, the methods currently used to determine the \fe\ velocity at day 50 must be improved, and spectral templates able to encompass the intrinsic variations of Type II-P SNe will be needed.
\end{abstract}


\section{Introduction}
Type Ia supernovae (SNe Ia) are standardized candles; although their peak magnitudes can vary by up to $\sim 2.5$ mag, empirical relations between the shape of their light curves and peak magnitude result in distance measurements that can be accurate to $\sim$7\% \citep{Phillips,JRK,Guy}.  This standardization has proven to be invaluable, as observations of SNe Ia led to the discovery of the accelerating expansion of the universe \citep{Riess98,Perl99}.  Dedicated surveys have now observed over a thousand of these objects in the past several years over a wide range of redshifts, lowering statistical uncertainties with the volume of their discoveries (For high redshift SN programs, see \citealp{Riess04,Riess07,astier06,WV07,miknaitis07,Kessler09,Frieman08}; for low redshift, see \citealp{LOSS,CSP,SNF,CFA}).  As a result, selection effects and other potential sources of bias are of increasing importance to improving the constraints on the dark energy equation of state.  In order to fully utilize future surveys that will discover very large numbers of SNe (DES\footnote{https://www.darkenergysurvey.org}; LSST\footnote{http://www.lsst.org/lsst}), these systematic uncertainties will have to be reduced.

Type II Plateau Supernovae (SNe II-P) can also be used as distance indicators, though thus far with less precision and to much lower distances than SNe Ia.  SNe II-P are objects that, observationally, are classified as Type II SNe by the presence of hydrogen absorption in their spectra, and as `Plateau' based on the slow decay of their early light curves \citep{barbon79}.  Distance measurements using the Expanding Photosphere Method \citep[EPM; see][]{EPM1,EPM2,EPM3} and the more recent, related Spectral Expanding Atmosphere Method \citep[SEAM; see][]{SEAM1,SEAM2}, which are based on the modeling of the SN atmosphere, have been shown to recover distances to 10\% precision \citep{SEAM3}.  A different approach has been taken by \citet[hereafter HP02]{HP02}, who have shown that SNe II-P magnitudes can also be standardized empirically, despite being far more heterogeneous \citep[peak luminosity dispersion of more than 5 mags; see][]{Patat,Pastorello} than SNe Ia.  The theoretically based methods take advantage of the relative simplicity of modeling hydrogen, which dominates the atmosphere of SNe II-P, as compared to the intermediate mass elements that make up the ejecta of SNe Ia.  SNe II-P also differ from their brighter brethren in a couple of other advantageous ways.  Over the past decade, observations in archival pre-SNe II-P images have found red supergiants in the mass range $~7-16\, M_{\odot}$ \citep{IIP1,IIP2,IIP3,IIP4,IIP5,IIP6,IIP7,IIP8}, and recently it has been shown that in at least one of these cases, the supposed progenitor is no longer present after the SN II-P has dimmed \citep{IIP9}.  So, compared to SNe Ia, where there is still uncertainty over the progenitor system (or systems), SNe II-P explosions are better understood.  Also, since SNe II-P have only been found in late-type galaxies - unlike SNe Ia, which are found in both late- and early-type galaxies - it is likely that biases from environmental effects will have a smaller effect on distance measurements for SNe II-P than SNe Ia.  Thus the differences between the two types of SNe and their methods of standardization will result in different systematic effects, allowing SNe II-P to serve as a useful check on SNe Ia measurements.

The standardization of the SNe II-P luminosity was introduced by HP02, who found that the luminosity and the expansion velocity at the photosphere are correlated when the SN is in its plateau phase; they use 50 rest-frame days post explosion as a convenient reference epoch to compare across SNe (this epoch is late enough to ensure the SN has entered the plateau phase, while still before it leaves the plateau).  The origin of this relationship is strongly grounded physically; a more luminous supernova's hydrogen recombination front will be at a greater distance from the core than in a less luminous SN, and thus the velocity of matter at the photosphere will be greater.  Using the velocity of the ejecta as measured from the minimum of the \feII\ absorption feature and a reddening correction based on the color of the SN when its light curve falls off the plateau, they found that the scatter in the Hubble diagram for \emph{V} and \emph{I} magnitudes drops from 0.95 and 0.80 mag, respectively, to 0.39 and 0.29 mag.  When they then restricted their sample to the 8 SNe with $z > 0.01$ in order to reduce the effect of peculiar velocities on host galaxy redshift, the scatter of their sample dropped even further, to 0.20 mag in \emph{V} and 0.21 mag in \emph{I}.  This technique has come to be known as the Standardized Candle Method (SCM).

\citet[hereafter N06]{N06} improved upon the HP02 method in several ways that address its limitations for application to SNe at cosmological distances.  The HP02 host galaxy extinction correction is replaced by the rest-frame \VmI\ color at day 50, which can be obtained with less extensive late time photometric monitoring.  To allow spectroscopic follow-up programs to be more flexible, they also derive an observational relationship between the velocity of the \fe\ line at epochs spanning days 9-75 to the velocity at day 50, which is crucial for any realistic spectroscopic follow-up program.  They apply the standardization method to 5 high-redshift ($0.13 < z < 0.29$) SNe obtained as part of the Supernova Legacy Survey \citep[SNLS;][]{astier06}, as well as one at $z = 0.019$ followed up by the Caltech Core-Collapse Program \citep[CCCP;][]{CCCP} (SN 1999gi, which was added to the dataset of SNe II-P by \citet{H03}, is also included).  The result of their SNLS-only fit differed strongly from that of the low-redshift SNe in HP02, which was attributed to a number of factors, including a small sample size, differences in data analysis techniques, and observational biases.  Their data did confirm, however, that the relation between ejecta velocity and magnitude could be used to reduce the intrinsic dispersion in the sample.

\citet{Olivares} applied the SCM to 37 SNe II-P, for which they use 30 days before the end of the plateau as the common epoch of reference.  They find the Hubble Diagram dispersion to be remarkably similar whether one uses \emph{B}-, \emph{V}-, or \emph{I}-band magnitudes (0.28, 0.31, and 0.32 mags, respectively), and raise the possibility that the magnitude-velocity relation may be quadratic instead of linear.

The most recent work on SNe II-P standardization was done by \citet[hereafter P09]{P09}, which presented new data on 19 low-redshift (all located at $z < 0.03$) SNe II-P with multi-epoch spectra.  These SNe were discovered as part of the Lick Observatory SN Search \citep[LOSS;][]{LOSS} by the Katzman Automatic Imaging Telescope (KAIT).  The most significant advancement of the SCM presented in P09 is the introduction of a robust technique for determining the velocity of the ejecta by cross-correlating the observed spectra with high-quality SNe II-P templates contained in SNID \citep[SuperNova Identification Code; see][]{SNID1,SNID2}.  As mentioned in P09, the velocity of the weak, broad \fe\ line is difficult to measure without introducing systematic offsets that hinder comparisons across multiple data sets.  Using the SCM, they found a significantly larger \emph{I}-band scatter (0.38 mag) in their Hubble Diagram than either HP02 or N06 (including all objects from both previous works), but showed that by removing all SNe with a positive plateau decline rate (defined as a decrease in brightness in \emph{I}-band magnitude between days 10 and 50) from their analysis, the dispersion of this `culled' sample was only 0.22 mag, or $\sim 10\%$ in distance.

In this work we apply the SCM to both the 15 new SNe II-P from the SDSS-II Supernova Survey and the combined set of our SNe plus those in the literature.  Our sample fills in a redshift range, $0.027 < z < 0.144$, that has very few other objects.  We discuss our observations in \S\ref{sec-obs}, and present light curves for these SNe and an additional 19 spectroscopically confirmed SNe II-P, also discovered during the SDSS-II Supernova Survey, for which we do not have sufficient data to include in our SCM sample.  We also discuss how we determine which SNe are part of the SCM sample.  In Section \S\ref{sec-phot} we explain in detail our process for determining K- and S-corrections, and explore how these are affected by the uncertainty in the explosion date and the choice of spectral template.  We analyze the spectra of our SNe in \S\ref{sec-spec}, and obtain their ejecta velocity at day 50.  In \S\ref{sec-cosm} we present the results from combining our observed SNe with those of HP02, N06, and P09, and applying the SCM to a combined sample of 49 SNe II-P.  We discuss our results and their interpretations in \S\ref{sec-results}, and conclude with an eye towards what will be needed for future SNe II-P campaigns in \S\ref{sec-dis}.  

\section{Observations}
\label{sec-obs}

The Sloan Digital Sky Survey-II Supernova Survey \citep[hereafter SDSS-SNS]{Frieman08} was one of three parts of the SDSS-II project.  The SDSS-SNS repeatedly surveyed the 300 square degree Southern Equatorial Stripe \citep[designated stripe 82; see][]{Stoughton} during the Fall seasons (September 1 - November 30) of 2005-2007 using a dedicated 2.5 meter telescope at Apache Point Observatory, New Mexico \citep{SDSS-Telescope}.  Each photometric observation consists of nearly simultaneous 55 second exposures in each of the five \emph{ugriz} filters \citep{SDSS-Filters} using the wide-field SDSS CCD camera \citep{SDSS-Camera}.  High quality light curves were obtained \citep{JH08} on a photometric system calibrated to an uncertainty of 1\% \citep{Ivezic}.  For a technical summary of the SDSS, see \citet{York}; further information can be found in \citet{Hogg}, \citet{Astrometry}, \citet{Photometric-System}, and \citet{Tucker}.

Supernova candidates were fit to model light curves of SNe Ia, SNe Ib/c, and SNe II, and photometrically classified by their most probable type; see \citet{Sako08}.  The main science driver of the SDSS-SNS was Type Ia SNe, so the spectroscopic follow-up program was designed to achieve high efficiency in following up these objects; only in the third season, when the work presented here began, did obtaining spectra of SNe II-P candidates become an explicit objective.  Spectroscopic observations of SNe II-P were performed by the 9.2m Hobby-Eberly Telescope (HET) at McDonald Observatory, 3.6m New Technologies Telescope (NTT), the 3.5m Astrophysical Research Consortium Telescope (ARC) at Apache Point Observatory, the 8.2m Subaru Telescope, the 2.4m Hiltner Telescope at MDM Observatory, the 4m Mayall Telescope at Kitt Peak National Observatory, the Magellan 6.5m Clay Telescope, the 4.2m William-Herschel Telescope (WHT), the 2.6m Nordic Optical Telescope (NOT), and the 3.6m Telescopio Nazionale Galileo (TNG).  Spectroscopic typing and redshift determination for each SN candidate is discussed in C. Zheng et al.(2009, in preparation); redshifts based on host galaxy spectra are accurate to $\Delta z = 0.0005$, and those based on SN spectral features are accurate to $\Delta z = 0.005$.

\subsection{SCM sample}
\label{sec-cuts}

The SDSS-SNS spectroscopically confirmed a total of 34 SNe II-P.  We do not include objects that are likely to be SNe II-P but have an unclear spectroscopic identity; this exclusion has no impact on our final sample of useable SNe.  To be classified as a SN II-P, the spectrum of the object must have identifiable hydrogen P-Cygni lines, and lack the narrow emission lines particular to the SN IIn class.  By finding all occasions where a SN candidate persists through more than one observing season, we eliminate from our sample any Active Galactic Nuclei (AGN).  Since no clear quantitative divide exists between the slowly declining SN II-P light curve and that of the more rapidly declining Linear class (SN II-L), we classify all of our remaining candidates as SNe II-P.  It is worth noting that, although we do not use the light curve shape to define our classification as a SN II-P, all objects classified this way exhibit flat or slowly declining linear light curves.

Our SCM sample is composed of the SDSS-SNS SNe II-P that pass three cuts, each of which removes only those SNe that lack the necessary data for application to the SCM; we do not remove any SN because of its observed properties.  The first cut, having an observational constraint on the explosion epoch, requires a null detection in the same observing season before the first detection of the SN (detection defined as 4$\sigma$ in flux).  This constraint eliminates all SNe II-P that are discovered in the first observation of each season.  The second cut imposed is on the length of the plateau observed.  To ensure an accurate derivation of the magnitude and color of each SN at day 50 (described in \S\ref{sec-phot}), 3 or more detections occurring at least 30 rest-frame days after the explosion - one of which must be at least 40 rest-frame days post-explosion - are required.  We impose this cut after comparing the day 50 flux interpolated from our most well sampled SNe with what we would obtain using only epochs at $t < 40$ to extrapolate the day 50 flux, and finding that the latter result can differ from the former by up to 50\%.  The photometry cut effectively excludes all SNe discovered roughly within the last half of each observing season.  The third requirement is that there be at least one spectrum taken between rest-frame days 9 and 75 that allows the \feII\ velocity at day 50 to be determined; details of this process are described in \S\ref{sec-spec}.  Since the nature of the spectroscopic follow-up program of the SDSS-SNS was primarily to determine the type of each supernova as early in its development as possible, the majority of our spectra are taken at early times (before the \fe\ line fully develops) and are of low signal-to-noise (since the objective of obtaining spectra was for typing, not analysis).  The main exception to the spectroscopic observing strategy was a single night of observing at Subaru during the Fall 2007 observing run (when SN II-P follow-up had become a priority) where we obtained high quality spectra for nine SNe II-P; six of these SNe passed our photometric cuts.  

\begin{figure}[htp]
  \centering
  \includegraphics[width=0.80\textwidth]{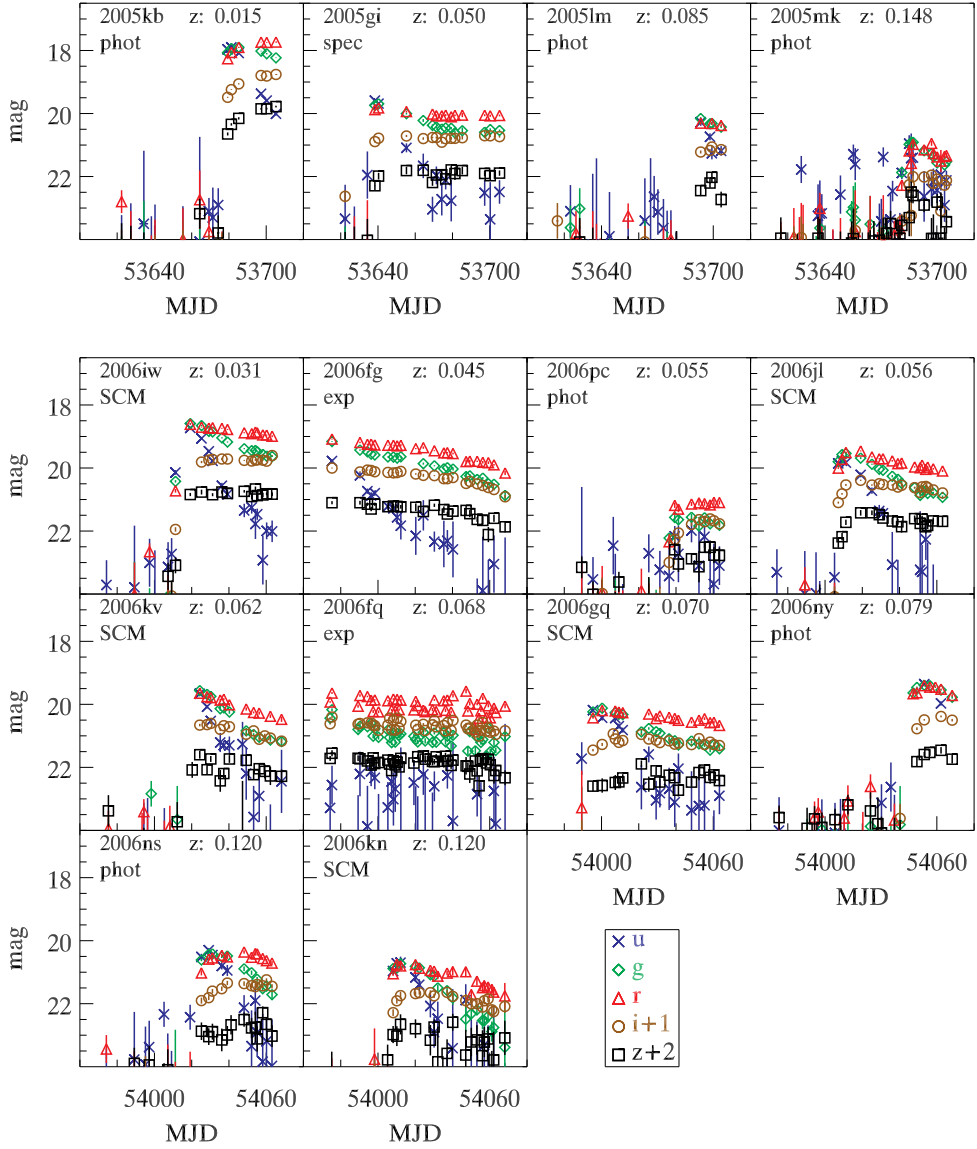}
  \caption{Observed light curves for SNe II-P from the 2005 (top) and 2006 (bottom) seasons of the SDSS-II Supernova Survey.
    Blue X's are magnitudes in \emph{u}-band, green diamonds are \emph{g}, red triangles are \emph{r}, brown circles are \emph{i}$+1$,
    and black squares are \emph{z}$+2$.  The x-axis is MJD-2400000.  The IAU name is given in the upper left of each box, and the
    redshift in the upper right.  Each SN that passes the cuts for inclusion in the Standard Candle Method sample, detailed in
    \S\ref{sec-cuts}, has the letters `SCM' printed beneath its redshift.  If the SN is not included in our SCM sample, it is 
    labeled as having failed the explosion epoch (exp), plateau photometry (phot), or spectroscopic (spec) cut, also described
    in \S\ref{sec-cuts}.  Only observations where the Scene Modeling Photometry (SMP; see \S\ref{sec-phot}) is deemed to be of 
    good quality are plotted; see \citet{JH08} for a description of SMP.  Magnitudes shown are asinh mags \citep{Lupton},
    with softening parameters given in \citet{Stoughton}.
    \label{fg-LC0506}}
\end{figure}

\begin{figure}[htp]
  \centering
  \includegraphics[width=0.80\textwidth]{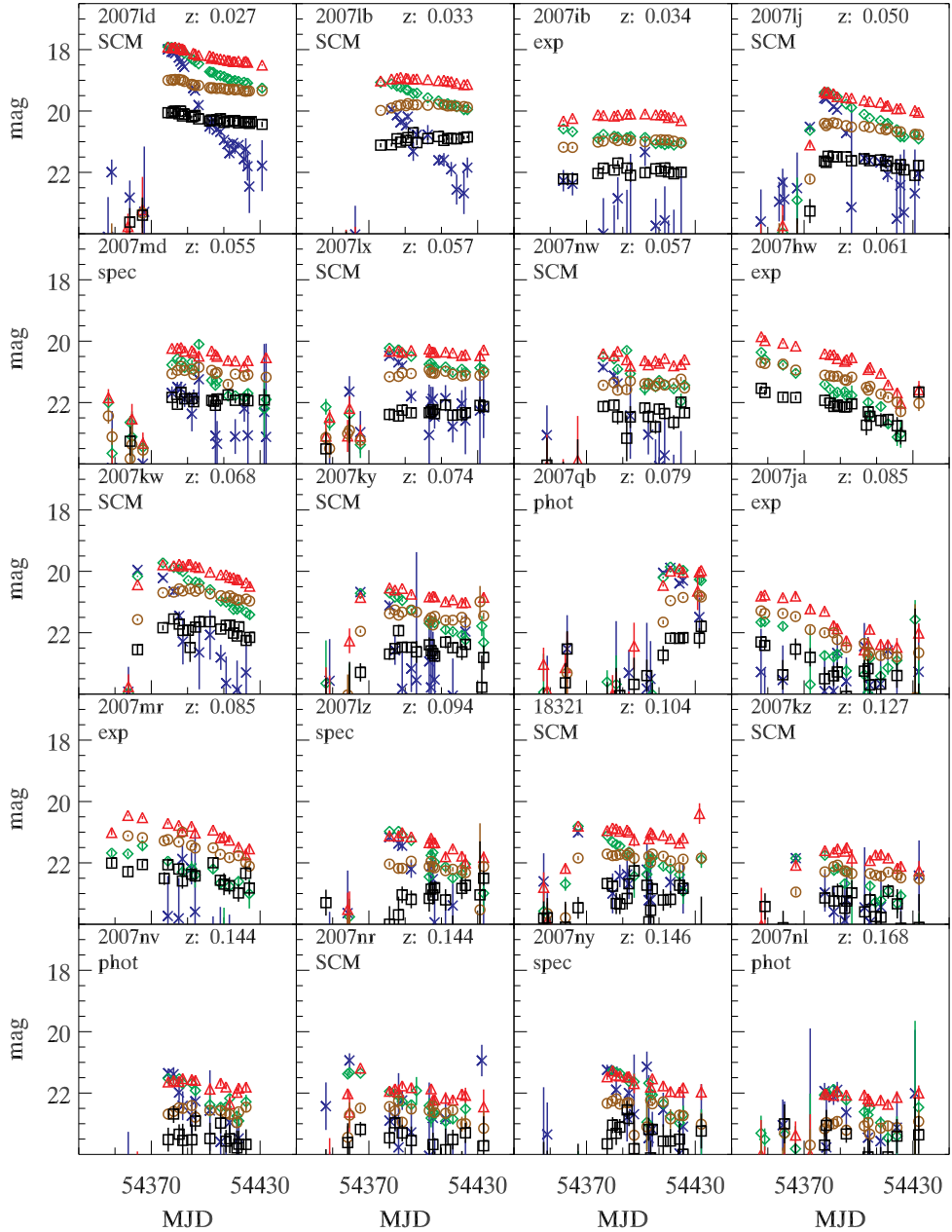}
  \caption{Same as Figure~\ref{fg-LC0506}, except for SNe discovered in the 2007 season.  Note the much larger number of SNe,
    which is primarily a result of a more concerted spectroscopic campaign for SNe II-P during the 2007 season.  The supernova 
    labeled `18321' is listed by its internal SDSS-SNS identification number, as it has not at present been assigned an IAU name.
    The two SNe that have rest-frame, day 50 \emph{I}-band uncertainties larger than 1 mag are 2007nl and 2007nv.
    \label{fg-LC07}}
\end{figure}

\begin{figure}[htp]
  \centering
  \includegraphics[width=0.84\textwidth]{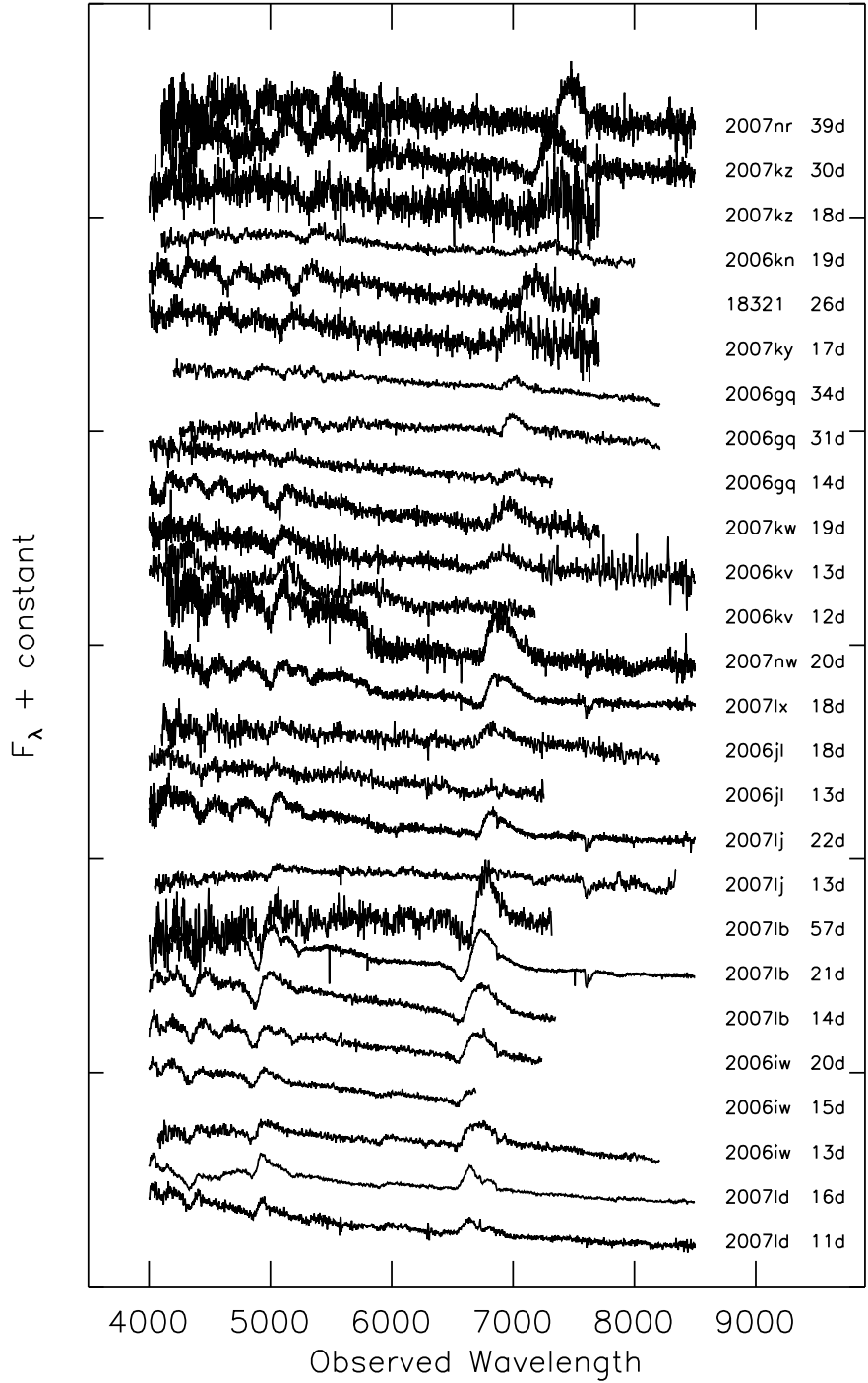}
  \caption{Spectra used in the SCM analysis for each of the 15 SNe II-P in our sample.  The spectra are shown in the observed frame, and the date listed for each SN is the number of days since explosion in the observer frame.  The redshift of each SN increases from the bottom ($z = 0.027$) to the top ($z = 0.1444$) of the plot.  We have masked galaxy emission lines in those cases where they were present.  Each spectrum has identifiable P-Cygni lines in H$\alpha$ and H$\beta$, although in some instances the weak \feII\ line cannot confidently be identified.  
    \label{fg-spec}}
\end{figure}

Applying these cuts to our observed SNe, we find 15 of the 34 spectroscopically confirmed SNe II-P have sufficient data for the SCM analysis, and thus form our SCM sample.  The explosion date constraint eliminates 6 objects; the photometric constraint 7; and the spectroscopic constraint 4.  Two more high-redshift SNe are eliminated from our sample after their rest-frame \emph{I}-band magnitudes are found to have an uncertainty greater than 1 mag.  We present the light curves for the 34 spectroscopically confirmed SNe II-P discovered by the SDSS-SNS in Figures \ref{fg-LC0506} and \ref{fg-LC07}.  Each of the 15 SNe that comprise the SCM sample as described above are labeled as such in these figures.  Approximately 60\% of the discovered SNe II-P are from the 2007 observing season; a change in the priority of these objects as spectroscopic targets is the reason for such a large percentage of our sample originating in one season.  All of the spectra that we use in our SCM analysis are plotted in Figure \ref{fg-spec}.  The spectra and photometry data for all 34 spectroscopically confirmed SNe II-P are available for public use through the SDSS-SNS website\footnote{http://sdssdp47.fnal.gov/sdsssn/photometry/SNIIP.tgz}, and are also included in the electronic version of this paper.

\clearpage

\section{Rest-Frame Plateau Magnitudes}
\label{sec-phot}

In this section we determine the rest-frame magnitudes at day 50 for each SN II-P in our sample.  First we determine the observed-frame magnitudes at day 50 through light curve fitting.  K-corrections are performed to transform observed magnitudes to the rest frame; S-corrections \citep{Stritzinger} are added to the \emph{ugriz} SDSS mags to obtain the Bessell \emph{BVRI} mags, which are required to compare our sample of SNe with those published by HP02, N06, and P09.  The following describes the procedure we adopted.

\begin{figure}[htp]
  \centering
  \includegraphics[width=0.55\textwidth]{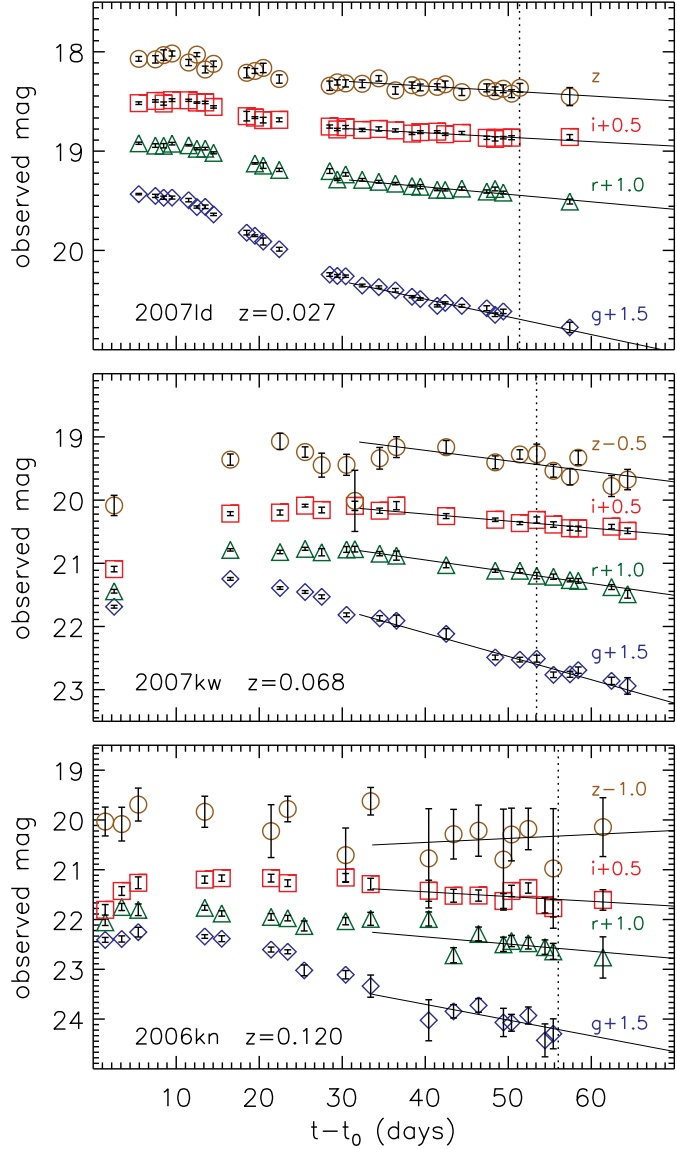}
  \caption{Observed SDSS-SNS light curves for SNe II-P at a range of redshifts.  Blue diamonds are \emph{g}-band mags; green triangles
    are \emph{r}-band mags; red squares are \emph{i}-band mags; and brown circles are \emph{z}-band mags.  For clarity, each light curve has been offset
    by an amount given on each plot.  Magnitudes have not been K- or S-corrected, nor have these light curves been corrected for 
    extinction.  Magnitudes are the standard definition of magnitudes, not the asinh mags defined in \citet{JH08} for SDSS SNe.  Solid
    lines are the plateau fits for each band, the procedure for which is defined in \S\ref{sec-phot}.  The dotted line denotes 
    rest-frame day 50 at the redshift of each SN.
    \label{fg-fitLCs}}
\end{figure}

The plateau of each SN is well described by a slow, linear evolution in magnitude; since we fit for the plateau in flux space, we model it as an exponential.  We fit the model to our observed photometry for all epochs $t-t_0 > 30(1+z)$, where $t_0$, the estimated time of explosion, is defined as the midpoint between the last pre-SN observation and the first detection epoch.  We use rest-frame day 30 as the starting point for the plateau because the observed rate of decline of the light curves in magnitudes after this epoch is constant.  We fit the plateau in the \emph{griz} filters only, excluding \emph{u} because its rapid decline precludes detections in this band for almost all of our SNe at late times; we show three examples of our plateau fits in Figure \ref{fg-fitLCs}.  Depending on how well-sampled each light curve is, we either interpolate or extrapolate fluxes at rest-frame day 50, and then convert them to SDSS filter magnitudes in the AB system using the zero-point offsets from the native SDSS system (see Table 1 of \citealp{JH08}).  We refer to these magnitudes throughout this paper as the day 50 magnitudes.

Next we transform the day 50 SDSS \emph{griz} mags into rest-frame Bessell \emph{BVRI} magnitudes.  For each SN, a template SN II-P spectrum is redshifted from its rest frame to the observed frame, and then reddened using the Milky Way extinction value along the line of sight from \citet{Schlegel}, the \citet[][hereafter CCM]{CCM} reddening law, and $R_V = 3.1$.  Since we do not have sufficient spectroscopic coverage of our SCM sample SNe to create our own day 50 SN II-P spectral template, we adopt the SN template used in N06 (P. Nugent 2009, private communication).  We then warp \citep{N02} the template spectrum to match the day 50 SDSS magnitudes using a spline; finally the template is de-extincted and returned to its rest-frame wavelength scale.  Rest-frame Bessell \emph{BVRI} magnitudes in the Vega system are then directly computed from the spectrum, using the response functions of \citet{MM} and the synthetic magnitudes of Vega in the Bessell filters given in Table 19 of \citet{Kessler09}.  The procedure described here serves to both K- and S-correct our observations.  We do not correct for host-galaxy extinction; since SNe II-P have similar colors along their plateau, the amount of host galaxy extinction is encapsulated in the rest-frame color.

Since there are uncertainties in the explosion epoch, the day 50 magnitudes, and the redshift of each SN, we run a Monte Carlo (MC) simulation of the process described above to determine how the uncertainties on these parameters affect the rest-frame day 50 \emph{BVRI} magnitudes.  Each iteration of the MC starts by selecting a date $t_{\mathrm{exp}}$ for the explosion epoch from the range $t_0 - \Delta t < t_{\mathrm{exp}} < t_0 + \Delta t$, where the uncertainty $\Delta t$ is half the difference between the last non-detection and the first detection of the SN and the probability distribution is described by a top hat function.  Due to the high cadence of the SDSS-SNS, the median uncertainty in explosion date from detection constraints in our SCM sample is an exceptionally low 3.5 days, matched only by the recent local sample presented in \citet{P09}.  The observed magnitude $m_x$ and statistical uncertainty $\sigma_{m_x}$ in each passband $x$ is taken from our plateau fit at day $t = t_{\mathrm{exp}} + 50(1+z)$, and is used to determine a realization of the observed-frame SDSS magnitudes at rest-frame day 50.  The uncertainty in the redshift is also included in the Monte Carlo, but as we have host-galaxy redshifts for most of our SNe, this source of uncertainty is small enough to have no significant effect on our results.  The rest-frame day 50 Bessell \emph{BVRI} magnitudes are then calculated for each iteration, from which we obtain the average magnitude in each passband and its uncertainty.  Although we cannot directly compare magnitudes in the SDSS filters to any other currently existing SN II-P dataset, we compute SDSS magnitudes (asinh mags are used by the SDSS-SNS, but apart from Figures \ref{fg-LC0506} and \ref{fg-LC07} we use log10 mags throughout this paper) and present them alongside the Bessell magnitudes in Table \ref{SDSS-data}, which represents the results for this section.

The uncertainty in the date of explosion does not significantly affect the determination of the day 50 magnitude in a SN II-P, since the plateau is modeled as a linear evolution and the uncertainty about the explosion date  is symmetric.  However, the same is not true for the uncertainty in the day 50 magnitude.  The plateau class, as discussed in P09, is not well defined, and the decline rate of SNe II-P can vary widely.  The greater the deviation from a completely flat plateau, the more important the explosion date uncertainty becomes.  By including the uncertainty in $t_0$ in our plateau modeling we obtain, on average, $\sim$10\% larger uncertainties in rest-frame \emph{I}-band magnitude and $\sim$60\% larger uncertainties in \emph{V}-band magnitude at day 50.  These magnitude uncertainties depend strongly on the slope of the plateau and the size of the explosion date uncertainty, and thus the size of the effect for each SN will vary.

\begin{deluxetable}{ll ccccc cccc}
\tablewidth{0pt}
\tabletypesize{\scriptsize}
\tablecaption{Rest-Frame Day 50 Magnitudes of SNe II-P from the SDSS-II SN Survey\label{SDSS-data}}
\tablehead{
  \colhead{IAU name} &
  \colhead{Redshift} & &
  \colhead{$B$} &
  \colhead{$V$} & 
  \colhead{$R$} &
  \colhead{$I$} & &
  \colhead{$g$} &
  \colhead{$r$} &
  \colhead{$i$} 
}
\startdata
2007ld  & 0.0270(0.0050)  & & 19.23(0.06) & 18.32(0.04) & 18.01(0.02) & 17.86(0.02) & & 18.77(0.06) & 18.15(0.02) & 18.22(0.02) \\
2006iw  & 0.0307(0.0005)  & & 19.82(0.02) & 18.97(0.01) & 18.56(0.01) & 18.31(0.02) & & 19.39(0.01) & 18.74(0.01) & 18.65(0.01) \\
2007lb  & 0.0330(0.0005)  & & 19.63(0.08) & 18.74(0.05) & 18.39(0.03) & 18.20(0.02) & & 19.18(0.07) & 18.54(0.03) & 18.55(0.02) \\
2007lj  & 0.0500(0.0050)  & & 20.79(0.06) & 19.90(0.04) & 19.55(0.03) & 19.37(0.04) & & 20.34(0.05) & 19.71(0.03) & 19.73(0.03) \\
2005gi  & 0.0505(0.0005)  & & 20.72(0.01) & 20.07(0.01) & 19.74(0.01) & 19.50(0.02) & & 20.35(0.01) & 19.92(0.01) & 19.84(0.01) \\
2007md  & 0.0546(0.0005)  & & 22.22(0.18) & 20.90(0.09) & 20.18(0.04) & 19.60(0.06) & & 21.65(0.15) & 20.43(0.05) & 20.12(0.04) \\
2006jl  & 0.0555(0.0005)  & & 20.50(0.03) & 19.65(0.02) & 19.32(0.01) & 19.10(0.04) & & 20.06(0.02) & 19.48(0.01) & 19.48(0.02) \\
2007lx  & 0.0568(0.0005)  & & 21.14(0.06) & 20.40(0.03) & 20.05(0.02) & 19.83(0.06) & & 20.75(0.05) & 20.22(0.03) & 20.17(0.03) \\
2007nw  & 0.0572(0.0005)  & & 21.65(0.10) & 20.63(0.04) & 20.26(0.03) & 19.92(0.10) & & 21.16(0.08) & 20.42(0.04) & 20.40(0.06) \\
2006kv  & 0.0620(0.0050)  & & 21.45(0.10) & 20.57(0.07) & 20.17(0.06) & 19.97(0.14) & & 21.01(0.08) & 20.34(0.06) & 20.31(0.08) \\
2007kw  & 0.0680(0.0005)  & & 21.24(0.06) & 20.28(0.03) & 19.86(0.02) & 19.57(0.03) & & 20.77(0.05) & 20.04(0.02) & 19.98(0.02) \\
2006gq  & 0.0698(0.0005)  & & 21.29(0.03) & 20.52(0.02) & 20.21(0.01) & 19.99(0.04) & & 20.88(0.03) & 20.37(0.01) & 20.36(0.02) \\
2007ky  & 0.0736(0.0005)  & & 22.05(0.07) & 21.01(0.03) & 20.56(0.02) & 20.22(0.06) & & 21.55(0.06) & 20.74(0.02) & 20.66(0.03) \\
2007lz  & 0.0940(0.0005)  & & 22.77(0.22) & 21.98(0.16) & 21.30(0.07) & 20.51(0.21) & & 22.38(0.19) & 21.62(0.10) & 21.10(0.12) \\
18321\tablenotemark{a}     &  0.1041(0.0005)  & & 22.01(0.08) & 21.19(0.04) & 20.83(0.03) & 20.73(0.09) & & 21.59(0.06) & 20.97(0.03) & 21.03(0.05) \\
2006kn  & 0.1203(0.0005)  & & 22.45(0.13) & 21.53(0.08) & 21.03(0.06) & 20.97(0.33) & & 22.02(0.10) & 21.19(0.07) & 21.21(0.16) \\
2007kz  & 0.1274(0.0005)  & & 22.71(0.10) & 21.81(0.06) & 21.27(0.04) & 21.25(0.28) & & 22.29(0.08) & 21.44(0.05) & 21.46(0.14) \\
2007nv  & 0.1439(0.0005)  & & 22.85(0.25) & 22.05(0.16) & 21.68(0.17) & 22.31(1.04) & & 22.47(0.21) & 21.77(0.17) & 22.24(0.55) \\
2007nr  & 0.1444(0.0005)  & & 22.67(0.09) & 22.17(0.06) & 21.72(0.05) & 21.78(0.29) & & 22.39(0.08) & 21.88(0.05) & 21.95(0.15) \\
2007ny  & 0.1462(0.0005)  & & 23.15(0.34) & 22.07(0.17) & 21.90(0.22) & 21.19(0.84) & & 22.57(0.23) & 22.09(0.22) & 21.96(0.45) \\
2007nl  & 0.1683(0.0005)  & & 22.14(0.36) & 22.03(0.26) & 21.66(0.41) & 20.57(1.29) & & 21.93(0.30) & 22.07(0.38) & 21.42(0.73) \\
\enddata 
\tablecomments{ Magnitudes listed here are derived for 50 days after the explosion, and have been both K-corrected and corrected for MW extinction along the line of sight.  Bessell \emph{BVRI} magnitudes are calculated in the Vega system and have been S-corrected from the observed filters; SDSS \emph{gri} magnitudes are standard log10 magnitudes (not asinh, as in \citealp{JH08}), and are in the AB system.  We model the plateau as an exponential in flux space, and fold into our photometric corrections the uncertainties in observed flux, explosion epoch and redshift.  The 21 SNe listed here include the 15 SCM sample SNe, as well as those that only fail our spectroscopy cut (described in \S\ref{sec-cuts}).}  
\tablenotetext{a}{We refer to SN18321 by its internal SDSS name, as it does not at the present time have an IAU designation.}
\end{deluxetable}

We implement two tests of our magnitude derivation procedure to ensure we have not induced an artificial bias into the SN magnitudes.  In the first test we attempt to reproduce the rest-frame day 50 magnitudes of the SNe presented in N06, the only other SNe II-P in the SCM literature that have been K- and S-corrected.  The 5 SNe II-P in N06 were observed by SNLS in the \emph{g'r'i'z'} bandpasses \citep{astier06}, and day 50 \emph{I}-band Vega-system magnitudes were derived.  We exclude from our comparison with the N06 analysis SNLS-03D3ce, which was not observed in the \emph{z'} bandpass, and SNLS-03D4cw, which had only 2 detections in the \emph{z'} bandpass beyond rest-frame day 30.  These two SNe do not pass our sample cuts (\S\ref{sec-cuts}), and therefore we are unable to correct to their day 50 magnitudes.  As previously stated, we use the same spectral template as in N06, so no systematic magnitude offset can be expected from this aspect of our warping method.  The observed fluxes and filter response functions used in the N06 analysis have been made publicly available by the authors\footnote{http://supernova.lbl.gov/$\sim$nugent/nugent\_papers.html}.  We find the difference of the rest-frame day 50 \emph{I}-band magnitudes derived in N06 from those we compute here to be 0.02 mag for SNLS-04D4fu, $-$0.03 mag for SNLS-04D1pj, and $-$0.05 mag for SNLS-04D1ln.  The good agreement with the N06 results, despite no description of the plateau-fitting method nor of the spectral warping process being described in their work, is taken as proof that our day 50 magnitudes are not biased relative to the SNe in N06.

As previously discussed, we fit a plateau to the observed frame magnitudes of our SNe II-P.  Since the Plateau class has classically been defined through observations of local SNe in \emph{BVRI} filters, it is reasonable to ask whether the day 50 magnitudes we obtain would be different if we were to fit a plateau to SNe that have already been K- and S-corrected.  Instead of performing our corrections on only the derived day 50 magnitudes (which we will call method A), we also correct each observation of each SN and fit a plateau to the resulting light curve (method B).  Since method B requires a set of finely spaced spectral templates, we interpolate for each epoch a spectrum from a set of SNe II-P templates\footnote{http://supernova.lbl.gov/$\sim$nugent/nugent\_templates.html}; the spectral template for the day 50 template in method A is taken from this interpolation as well.  To account for the uncertainty in explosion epoch in method B, we construct distinct rest-frame light curves that assume for the true explosion epoch each integral value in the range $t_0 - \Delta t < t_{\mathrm{exp}} < t_0 + \Delta t$.  A plateau is fit to each of these light curves, and assuming an equal likelihood for each value of $t_{\mathrm{exp}}$, a MC is used to determine the flux at day 50.  The average difference in day 50 magnitudes derived by these two methods is $0.006 \pm 0.008$ mags.  Since both methods give the same results, we use method A throughout this paper because it is quicker to run and does not require a full time series of spectral templates.

\section{Spectroscopic Analysis}
\label{sec-spec}

The expansion velocity of the ejecta in SNe II-P is the key to the SCM, as it is this parameter that correlates with luminosity; over-luminous SNe have high expansion velocities and under-luminous SNe have low velocities (HP02).  The ejecta velocity is determined by measurements of the weak \fe\ absorption lines around $\lambda \approx 5000$\AA, in particular the \feII\ line \citep{H03}.  Unfortunately, the velocity of this feature is often difficult to measure.  Two methods have been explored in previous studies for extracting the \fe\ velocity from noisy spectra.  The  H$\beta$ absorption feature, which is often stronger (particularly at early times) than the \fe\ feature, is shown in N06 to be related to the \fe\ velocity in such a way that the measurement of the H$\beta$ velocity can recover the former to $\pm$ 300 km~s$^{-1}$.  A different approach is taken in P09, where the \fe\ velocity is determined through cross-correlating observed spectra with a library of SNe II-P using SNID (this method is explored in N06, but more thoroughly developed in P09).  Each spectrum in the library has its \fe\ velocity measured from the absorption-line minimum, and a weighted mean is taken of all velocities measured from good spectral fits (See P09 for more detail).  We use this cross-correlation method to derive the \fe\ velocities from our SNe, as it has several advantages:  it is an easily reproducible method; uncertainty due to low signal-to-noise ratio (S/N) spectra is incorporated through the goodness of fit of the correlations; and it allows for a straightforward comparison with the full sample of SNe II-P compiled by P09.  We list in Table \ref{SDSS-vel} the velocities and statistical uncertainties derived using this method for each spectrum of a SN II-P in our SCM sample (shown in Figure \ref{fg-spec}).

After computing the \fe\ velocity of each spectrum with the above technique, we extrapolate the velocity of the ejecta at $t = 50$ days using Equation 1 from N06,
\begin{equation}
\label{eq-vel}
v_{\mathrm{Fe\,II}}(50) = v_{\mathrm{Fe\,II}}(t)(t/50)^{0.464 \pm 0.017}.
\end{equation}
The range of epochs that this function, which is derived from fitting published velocities of SNe II-P with multi-epoch spectra, is claimed to be valid for is days $9$-$75$ post-explosion.  We only list in Table \ref{SDSS-vel} spectra that fall in this range; we leave out our spectra that were taken at earlier epochs.  The day 50 ejecta velocities for our SCM sample of SNe II-P are shown in Table \ref{SDSS-SCM}.  For those SNe that have multi-epoch spectra we have taken a weighted average of the velocity derived from each spectrum.  The large uncertainties in velocity for some SNe are a result of either having only one spectrum at an early time, or having a larger uncertainty ($> 5$ days) in the explosion epoch; for several SNe both conditions apply.  The uncertainties in \vel(50) given in Table \ref{SDSS-SCM} do not include the effects of peculiar motion, which contributes an uncertainty in velocity of 300 km~s$^{-1}$.  We add this uncertainty in quadrature to that in Table \ref{SDSS-SCM} when we use our data in the SCM.

\begin{deluxetable}{lllc}
  \tablewidth{0pt}
  \tabletypesize{\scriptsize}
  \tablecaption{Observed \fe\ velocities of SNe II-P from the SDSS-II SN Survey\label{SDSS-vel}}
  \tablehead{
    &
    \colhead{$t - t_0$\tablenotemark{a}} &
    \colhead{\vel} &
    \\
    \colhead{IAU name} &
    (days) &
    (km~s$^{-1}$) &
    \colhead{Observatory}
  }
  \startdata
2007ld &  11(5.5)  &  7860(720)  &  APO \\
       &  16(5.5)  &  7370(710)  &  NTT \\
2006iw &  13(1.0)  &  7870(750)  &  NTT \\
       &  15(1.0)  &  8090(580)  &  MDM \\
       &  20(1.0)  &  7040(240)  &  MDM \\
2007lb &  14(7.0)  &  6830(520)  &  MDM \\
       &  22(7.0)  &  5700(400)  &  Subaru \\
       &  59(7.0)  &  3870(430)  &  MDM \\
2007lj &  22(3.5)  &  6310(560)  &  Subaru \\
2006jl &  13(1.0)  &  11700(850) &  MDM \\
       &  18(1.0)  &  8630(590)  &  NTT \\
2007lx &  18(8.0)  &  6430(220)\tablenotemark{b}  &  Subaru \\
2007nw &  20(7.0)  &  6730(440)  &  Subaru \\
2006kv &  13(4.0)  &  7000(610)  &  APO \\
2007kw &  19(2.5)  &  6350(670)  &  TNG \\
2006gq &  14(3.0)  &  5680(530)  &  MDM \\
       &  31(3.0)  &  4280(210)  &  NTT \\
       &  34(3.0)  &  4430(210)  &  NTT \\
2007ky &  17(3.0)  &  5990(500)  &  TNG \\
18321  &  26(5.0)  &  7170(630)  &  TNG \\
2006kn &  19(1.5)  &  6970(700)  &  NTT \\
2007kz &  18(3.5)  &  6660(400)  &  TNG \\
       &  30(3.5)  &  7150(680)  &  Subaru \\
2007nr &  39(5.0)  &  5290(370)  &  Subaru \\
\enddata 
\tablecomments{  \,Listed are all of the spectra taken between rest-frame days 9 and 75 after the explosion for 
our SCM sample; note the bias in observed epoch towards early times.  The velocity of the \fe\ line and its 
uncertainty for each spectrum is obtained from the cross-correlation code of P09.  Uncertainties in explosion epoch
and observed-day velocity are given in parentheses.}
\tablenotetext{a}{Observer-frame post-explosion epoch at which each spectrum was obtained.}
\tablenotetext{b}{We remove galaxy emission contamination before computing \vel\ for SN 2007lx, as there is a strong
  emission line next to the \feII\ absorption feature that biases the cross-correlation fit; without masking the galaxy 
  lines we measure 6010(400) km~s$^{-1}$.  We do not use masked spectra for any of the other \vel\ measurements in 
  this table, as the galaxy lines for these spectra have no notable effect.}
\end{deluxetable}

\clearpage

\section{Standardized Candle Method}
\label{sec-cosm}

The Standardized Candle Method assumes the magnitudes of SNe II-P are correlated with the velocity of the ejecta, as measured by the \feII\ absorption line.  The observed magnitude in the rest-frame \emph{I}-band has been shown empirically to obey the relation
\begin{equation}
\label{eq-fit}
m_I = {\cal M}_{I_0} - \alpha \log_{10}(v_{\mathrm{Fe\,II}}/5000) + R_I[(V-I)-(V-I)_0] + 5 \log_{10}({\cal D}_L(z|\Omega_M,\Omega_{\Lambda})),
\end{equation}
where the K-corrected apparent magnitude $m_I$, color \VmI\, absolute magnitude ${\cal M}_{I_0}$, and \fe\ velocity \vel\ (in units of km~s$^{-1}$) are all measured at 50 days post explosion.  We follow HP02, N06, and P09 in performing the standardization technique using the \emph{I}-band magnitude.  As in P09, we allow the extinction law $R_I \equiv A_I/E(V-I)$ to be a free parameter in our analysis, and adopt the ridge-line SN II-P color from N06, $(V-I)_0 = 0.53$ mag (the exact value of this color is not important, as the parameter is degenerate with ${\cal M}_{I_0}$).  Since we are measuring only relative distances, we use the ``Hubble Constant free'' absolute magnitude and luminosity distance, ${\cal M}_{I_0} \equiv M_{I_0} - 5\log_{10}(H_0) + 25$ and ${\cal D}_L \equiv H_0 D_L$ \citep[P09;][]{Goobar95,Sullivan06}.  We assume throughout this paper a fiducial $\Lambda$CDM cosmology with $\Omega_{M} = 0.3$, since we are unable at these redshifts to put a meaningful constraint on cosmology.  Though we fit for ${\cal M}_{I_0}$, we quote values of $M_{I_0}$ throughout this paper, assuming $H_0 =$ 70~km~s$^{-1}$~Mpc$^{-1}$.  This allows for a more intuitive understanding of the standard absolute magnitude of SNe II-P.

We find the best fit values of our free parameters by running a Monte Carlo Markov Chain (MCMC) to minimize the negative log of the likelihood function with respect to the velocity term coefficient $\alpha$; the color law $R_I$; the ``Hubble Constant free'' absolute magnitude ${\cal{M}}_{I_0}$; and the intrinsic dispersion $\sigma_{\mathrm{int}}$.  Correcting an error in equation 4 of P09, we have 
\begin{equation}
\label{eq-like}
-2\log({\cal{L}}) = \sum_{SN}\left\{\,\frac{\left[m^{\mathrm{obs}}_I- m_I(v_{\mathrm{Fe\,II}},V-I,z;\, {\cal{M}}_{I_0},\alpha,R_I)\right]^2}{\sigma^2_{\mathrm{total}}} + \log(\sigma^2_{\mathrm{total}})\right\},
\end{equation}
where the summation is over all SNe in the sample; $m_I(v_{\mathrm{Fe\,II}},V-I,z;\, {\cal{M}}_{I_0},\alpha,R_I)$ is defined by equation \ref{eq-fit}; and the total uncertainty is
\begin{equation}
\sigma^2_{\mathrm{total}} = \sigma^2_{m_I}+\left(\frac{\alpha}{\ln 10}\,\frac{\sigma_{v_{\mathrm{Fe\,II}}}}{v_{\mathrm{Fe\,II}}}\right)^2+\left(R_I\,\sigma_{(V-I)}\right)^2+\left(\frac{5}{\ln 10}\,\frac{\sigma_{{\cal{D}}_L(z)}}{{\cal{D}}_L(z)}\right)^2+\sigma^2_{\mathrm{int}}.
\label{eq-sigma}
\end{equation}
The logarithmic term in equation \ref{eq-like} comes from the normalization of the likelihood function; this term weights against excessively large values of the free parameters, which could otherwise be favored by the first part of the log-likelihood function.  The variable $\sigma_{\mathrm{int}}$ represents the amount of intrinsic scatter among SNe II-P that cannot be accounted for by the model itself, and thus represents the minimum statistical uncertainty in any distance determination from the SCM.

In Table \ref{SDSS-SCM} we present the rest-frame magnitudes and velocities at day 50 for the 15 SNe II-P that comprise the SCM sample from the SDSS-SNS.  When our SNe are added to the existing sample of SNe II-P with measurements at day 50 (presented in Table 2 of P09), the SCM can be applied over a nearly continuous redshift range, $0.001 < z < 0.25$.

\begin{deluxetable}{lcccc}
\tablewidth{0pt}
\tabletypesize{\scriptsize}
\tablecaption{Parameters for SCM on SNe II-P from the SDSS-II SN Survey \label{SDSS-SCM}}
\tablehead{
  &
  &
  \colhead{\vel} &
  &
  \\
  \colhead{IAU name} &
  \colhead{Redshift} &
  \colhead{(km~s$^{-1}$)} &
  \colhead{\emph{I}} & 
  \colhead{\VmI\ }
}
\startdata
2007ld &  0.0270(0.0050)  &   4105(574)  &  17.86(0.02)  &  0.46(0.04)\\
2006iw &  0.0307(0.0005)  &   4488(164)  &  18.31(0.02)  &  0.66(0.02)\\
2007lb &  0.0330(0.0005)  &   3945(337)  &  18.20(0.02)  &  0.54(0.05)\\
2007lj &  0.0500(0.0050)  &   4215(464)  &  19.37(0.04)  &  0.53(0.06)\\
2006jl &  0.0555(0.0005)  &   5560(313)  &  19.10(0.04)  &  0.55(0.04)\\
2007lx &  0.0568(0.0005)  &   3901(819)  &  19.83(0.06)  &  0.57(0.07)\\
2007nw &  0.0572(0.0005)  &   4287(754)  &  19.92(0.10)  &  0.71(0.11)\\
2006kv &  0.0620(0.0050)  &   3644(616)  &  19.97(0.14)  &  0.60(0.16)\\
2007kw &  0.0680(0.0005)  &   3931(462)  &  19.57(0.03)  &  0.71(0.04)\\
2006gq &  0.0698(0.0005)  &   3404(149)  &  19.99(0.04)  &  0.53(0.04)\\
2007ky &  0.0736(0.0005)  &   3513(417)  &  20.22(0.06)  &  0.79(0.07)\\
18321  &  0.1041(0.0005)  &   5056(636)  &  20.73(0.09)  &  0.46(0.10)\\
2006kn &  0.1203(0.0005)  &   4220(443)  &  20.97(0.33)  &  0.56(0.34)\\
2007kz &  0.1274(0.0005)  &   4376(322)  &  21.25(0.28)  &  0.56(0.29)\\
2007nr &  0.1444(0.0005)  &   4428(349)  &  21.78(0.29)  &  0.39(0.30)\\
\enddata 
\tablecomments{Derived SN parameters required for the SCM analysis.  Magnitudes and velocities 
  listed are for $t = t_0 + 50(1+z)$; the derivation of these values are described in \S\ref{sec-phot}
  and \S\ref{sec-spec}.  Magnitudes have been K- and S-corrected, and corrected for Milky Way extinction.
  An additional uncertainty of $\sigma_{\mathrm{pec}} = 300$ km~s$^{-1}$ is added in quadrature to both the 
  redshift $z$ and the velocity derived from the \feII\ absorption feature, \vel, before applying the SCM
  to this sample.}
\end{deluxetable}

\clearpage

\section{Results}
\label{sec-results}

\begin{figure}[htp]
  \centering
  \includegraphics[width=0.8\textwidth]{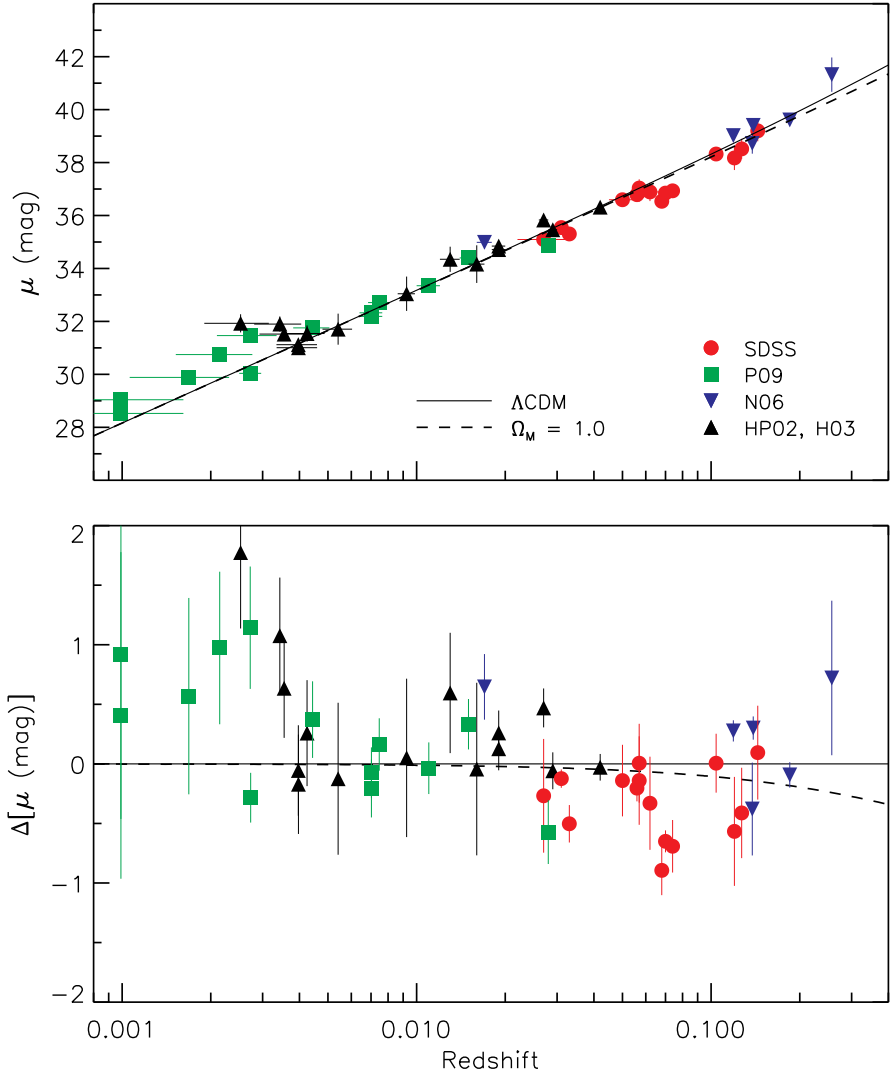}
  \caption{Top panel:  Hubble Diagram for SNe II-P, using the SCM and parameters derived in this work.  Black triangles are data from HP02 and H03; green squares are from P09; blue upside-down triangles are high-$z$ SNe from N06; and red circles are from this paper.  As an example of the effects of cosmology in the redshift range covered by the data in this work, we plot the Hubble Law for $\Lambda$CDM, $\Omega_{M} = 0.3$ (solid line) and $\Omega_{M} = 1$ (dashed line) cosmologies.  The distance modulus for each SNe is defined as $\mu = m^{\mathrm{obs}}_I - (M_{I_0} - \alpha \log_{10}(v_{\mathrm{Fe\,II}}/5000) + R_I[(V-I)-(V-I)_0])$, and the Hubble constant is assumed to be 70~km~s$^{-1}$~Mpc$^{-1}$.  Bottom Panel:  Residuals from the Hubble Diagram.  We project uncertainties in redshift onto the distance modulus using the distance-redshift relation, $\sigma_{\mu} = \sigma_{{\cal{D}}_L(z)}(5/\ln 10)$.
    \label{fig-Hubble}}
\end{figure}

Combining the 15 SDSS SNe II-P in Table \ref{SDSS-SCM} with the 34 SNe II-P in the `culled' set of P09, the best-fit SCM parameters as determined by minimizing Equation \ref{eq-like} are $M_I = -17.52 \pm 0.08$, $\alpha = 4.0 \pm 0.7$, and $R_I = 0.8 \pm 0.3$, with an intrinsic dispersion $\sigma_{\mathrm{int}} = 0.29$ mag.  The Hubble Diagram and the Hubble residuals of the combined data set is plotted in Figure \ref{fig-Hubble}.  The scatter found here is comparable to the results of N06 and \citet{Olivares}, but the addition of the SDSS SNe increases the dispersion from the 0.22 mag found in P09.

To find the dispersion of SNe II-P from a true standard candle, we minimize Equation \ref{eq-like} while fixing the parameters $\alpha$ and $R_I$ to zero.  HP02 quote an \emph{I}-band dispersion without any corrections of 0.80 mag, and P09 state the dispersion of the `culled' sample to be $\approx$ 0.9 mag.  When we solve for the dispersion from a true standard candle of the P09 `culled' sample using our Equation \ref{eq-like}, we find $\sigma_{\mathrm{int}} = 0.67$ mag.  The SDSS-only set of SNe II-P introduced here has a much smaller true intrinsic dispersion; $\sigma_{\mathrm{int}} = 0.22$ mag.  We show in \S\ref{sec-selection} that this tight scatter can be attributed to selection effects.

\begin{figure}[htp]
  \centering
  \includegraphics[width=0.8\textwidth]{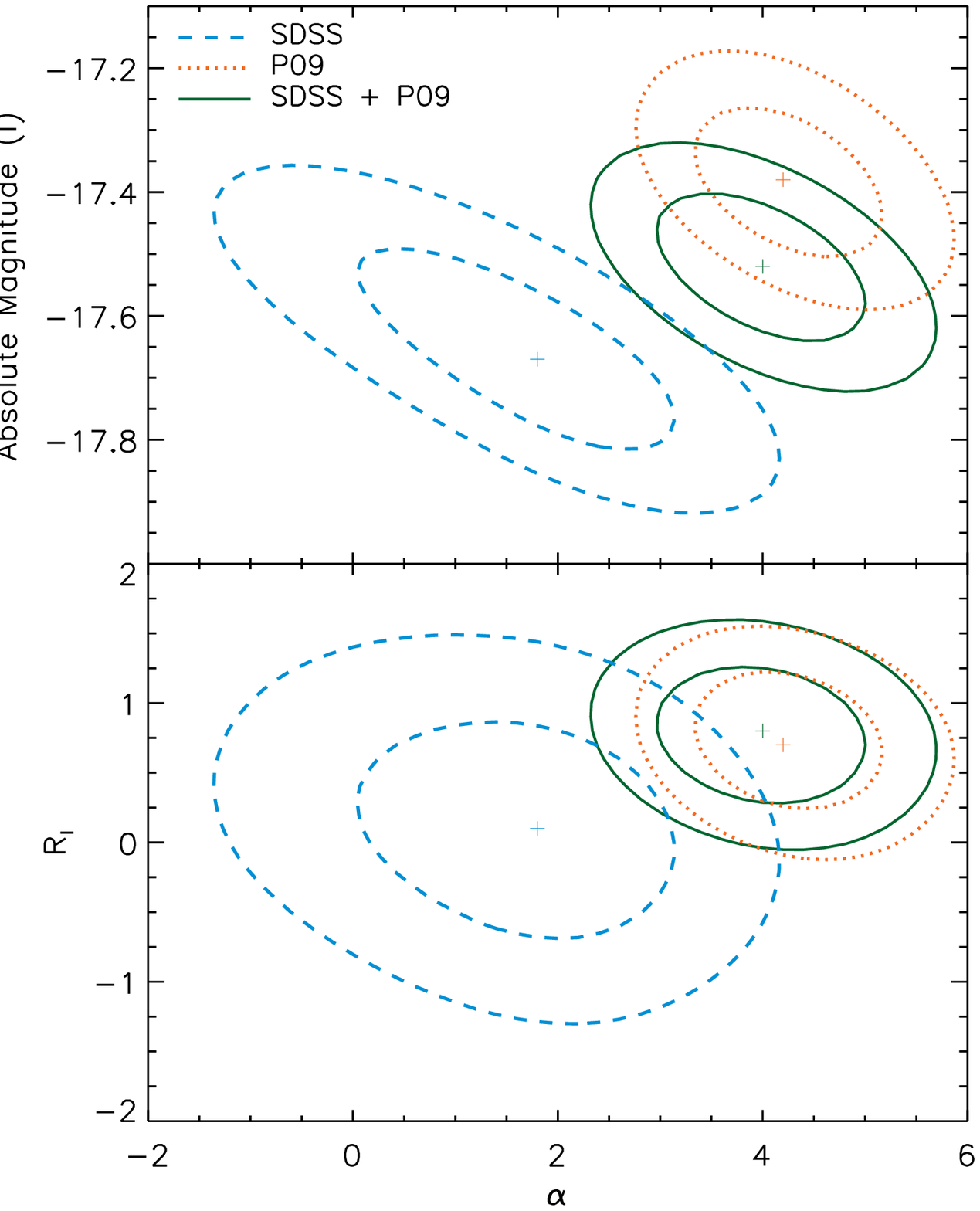}
  \caption{The marginalized 1$\sigma$ and 2$\sigma$ confidence intervals in the $\alpha$ - $M_{I_0}$ (Top) and $\alpha$ - $R_I$ (Bottom) planes are shown here.  The orange dotted line is for the P09 `culled' sample; the blue dashed line is the SDSS only sample; and the green solid line is the joint set.  
\label{fig-contour}}
\end{figure}

Using the SCM reduces the intrinsic scatter in the SDSS SNe II-P from 0.22 to 0.16 mag, with best fit parameters of $M_I = -17.67^{+0.11}_{-0.10}$, $\alpha = 1.8^{+0.9}_{-1.0}$, and $R_I = 0.1 \pm 0.5$.  These results are significantly different than the `culled' sample of 34 SNe in P09, whose parameters we calculate to be $M_I = -17.38 \pm 0.08$, $\alpha = 4.2 \pm 0.6$, and $R_I = 0.8 \pm 0.3$.   The SDSS SNe have a standard magnitude that is brighter than the pre-existing sample by $> 2 \sigma$; have a weaker correlation with the velocity of the ejecta; and display no strong correlation between color and luminosity.  Figure \ref{fig-contour} projects the likelihood surface of the two samples onto the $\alpha$ - $M_{I_0}$ and $\alpha$ - $R_I$ planes, showing that in the former these two samples do not intersect at the 2$\sigma$ level.  It is interesting to note that the high-redshift SNe of N06 also did not intersect with the low-redshift SNe of HP02 at the 2$\sigma$ level in the $\alpha$ - $M_{I_0}$ plane (see Figure 14 of N06).

We also note that the SCM parameters we give for the P09 `culled' set are not the same as the published values.  These parameter differences exist because our Equations \ref{eq-like} and \ref{eq-sigma} are slightly different from what was used in P09.  We stress that the differences we find are small; $\alpha$ is lower by 0.2, $R_I$ is unchanged, $M_I$ is 0.01 mag more luminous, and the intrinsic dispersion, 0.20 mag, is 0.02 smaller than in P09.  The SCM parameters we list in Table \ref{SCM-fits} for various SNe II-P sample combinations and those stated throughout the remainder of the paper are, unless stated otherwise, calculated using the equations in \S\ref{sec-cosm}.

In P09, the initial result for $\sigma_{\mathrm{int}}$ with their full sample of 40 SNe II-P was 0.38 mag.  The three SNe with the largest Hubble residuals (in units of standard deviations) in P09 all were found to have positive decline rates in the \emph{I}-band between days 10 and 50 post-explosion.  By removing all SNe from their sample that had a positive decline rate at the 1$\sigma$ level, P09 obtained a value for $\sigma_{\mathrm{int}}$ of only 0.22 mag.  We calculate the intrinsic scatter of the `full' and `culled' sets in Table \ref{SCM-fits} to be 0.35 and 0.20 mag, respectively.  In total, 6 of their 40 SNe, all of which came from the 19 SNe detected by KAIT, were eliminated with the decline rate cut.

To find the slope of the \emph{I}-band light curve for every SDSS SN II-P, we K- and S-correct each observed epoch in the light curve of each SN; this process is described as method B in \S\ref{sec-phot}.  We find that the rest-frame \emph{I}-band light curve has a positive decline rate for 6 of our 15 SNe.  Plotting decline rate against the distance of each SDSS SN II-P from the Hubble line derived from the SDSS-plus-P09 sample (in units of standard deviation), we find that none of the five most deviant SNe in our sample would be removed using the decline rate method.  Therefore we cannot use the culling method of P09 to identify a more standard subclass of SNe II-P.  Whereas the inclusion of the SDSS SNe II-P substantially increases the intrinsic \emph{I}-band dispersion of the `culled' P09 sample, the intrinsic dispersion of the full sample is unchanged by this addition.

We also comment here on the effect of peculiar velocity on our results.  We state in \S\ref{sec-spec} that for our SNe II-P sample we add in quadrature 300~km~s$^{-1}$ to the uncertainty in \vel\ to account for the peculiar velocity of the SNe within its host galaxy.  In P09 a smaller uncertainty is added to the expansion velocity of each SN, and this source of uncertainty is not mentioned at all in N06.  If we change the uncertainty of each SN in the P09 `culled' and `full' sets to reflect a $\sigma_{v_{\mathrm{Fe\,II}},\mathrm{pec}}$ of 300~km~s$^{-1}$, we find no significant difference in the best-fit SCM parameters of the combined data sets.  The intrinsic scatter of the P09 SNe would be smaller if this peculiar velocity uncertainty were used, but the parameters again would remain relatively unchanged.  Since our main conclusions are not affected by this source of uncertainty, we choose to adopt the \vel\ uncertainties as is from Table 2 of P09.

\begin{deluxetable}{l|cccc}
\tablewidth{0pt}
\tabletypesize{\scriptsize}
\tablecaption{SCM Fit Parameters \label{SCM-fits}}
\tablehead{
  &
  \colhead{$\alpha$} &
  \colhead{$R_I$} &
  \colhead{$M_{I_0}$\tablenotemark{a}} &
  \colhead{$\sigma_{\mathrm{int}}$}
}
\startdata
SDSS          &  1.8($^{+0.9}_{-1.0}$)  &  0.1(0.5)               &  -17.67($^{+0.11}_{-0.10}$)  &  0.16(0.07)                 \\
P09 `culled'  &  4.2(0.6)               &  0.8(0.3)               &  -17.38(0.08)                &  0.20(0.07)                 \\
P09 `full'    &  4.4(0.7)               &  0.6($^{+0.3}_{-0.4}$)  &  -17.42($^{+0.11}_{-0.10}$)  &  0.35($^{+0.08}_{-0.07}$))  \\

SDSS + P09 `culled' & 4.0(0.7)          &  0.8(0.3)               &  -17.52(0.08)                &  0.29(0.06)                 \\
SDSS + P09 `full'   & 4.0(0.7)          &  0.7(0.3)               &  -17.53(0.09)                &  0.35(0.06)                 \\

\enddata 
\tablecomments{Best-fit values and 1$\sigma$ uncertainties for each parameter in the SCM fit, marginalized over the other parameters.}
\tablenotetext{a}{The standard absolute \emph{I}-band magnitude $M_{I_0}$ is derived from ${\cal M}_{I_0}$ assuming $H_0 = $\,70~km~s$^{-1}$~Mpc$^{-1}$.}
\end{deluxetable}
\clearpage

\subsection{Selection Effects and Biases}
\label{sec-selection}

The disparity between the SCM fit parameters from the SDSS SNe and those presented in P09 requires explanation.  We take a close look at selection effects, as well as the techniques used to obtain the magnitudes and velocities in Table \ref{SDSS-SCM}, for potential sources of bias that could cause our results to differ from previous works.

\begin{figure}[htp]
  \centering
  \includegraphics[width=0.6\textwidth]{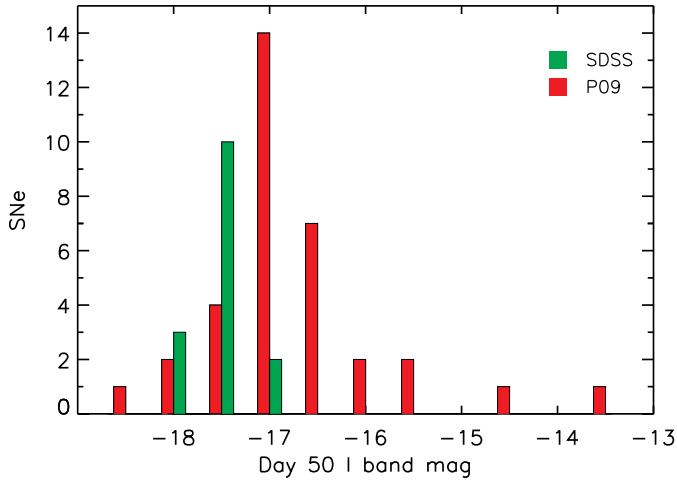}
  \caption{Histogram showing the absolute \emph{I}-band magnitudes of SNe II-P observed as part of the SDSS-SNS (Green) and previous studies (`Culled' sample in P09; Red).  Magnitudes are calculated assuming a Hubble Constant of 70~km~s$^{-1}$~Mpc$^{-1}$, and $\Omega_{\Lambda} = 0.7, \Omega_{M} = 0.3$.  The SCM sample we construct from the SDSS-SNS is strongly biased against the inclusion of faint SNe II-P.
\label{fig-absmags}}
\end{figure}

One possible explanation of the relatively high luminosity of our sample when compared with previous SNe II-P studies is selection effects (though selection effects may have affected the results of previous work as well, as mentioned in N06).  To examine the effects of Malmquist bias, we calculate the absolute magnitudes of the SDSS SNe II-P assuming our fiducial cosmology.  The narrow range of day 50 magnitudes we find, $-17.1 < M_I <  -17.9$, conflicts with the well known fact that SNe II-P occupy a range of luminosities spanning at least two orders of magnitude (See Figure \ref{fig-absmags}).  Since the data presented here are a compilation of objects discovered by the SDSS-SNS during drift-scan searches, there is no expectation of completeness amongst the data, and a bias towards intrinsically brighter objects above some redshift is inevitable.  We use the Supernova Analysis package \citep[SNANA;][]{SNANA} and perform an analysis similar to that in \citet{Dilday08,Dilday09} to determine the probability of detecting a SN II-P of a given luminosity in the SDSS-SNS.  We find that for a SN II-P with an absolute \emph{r}-band magnitude of $-$16.0 in its plateau (absolute \emph{I}-band magnitude of $-$16.54) located at a redshift of 0.07 (apparent \emph{r}-band magnitude of 21.63), the detection efficiency by the SDSS-SNS is $\sim 75$\%.  At a redshift of 0.054, the probability of detection is $\sim 90$\%.  This effect clearly explains why our SNe II-P sample does not contain any dim SNe above $z \approx 0.10$.

While Malmquist bias limits the magnitude distribution of our sample at higher redshift, it does not explain why we do not see intrinsically dimmer SNe at lower redshifts.  However, not all supernova candidates discovered in the survey were observed spectroscopically, which is for obvious reasons a necessary condition for the SNe in our SCM sample.  While candidates with a peak observed \emph{r}-band magnitude of $\approx$ 20.5 or brighter were spectroscopically followed up in all cases, dimmer candidates were subject to more selective criteria \citep{Sako08}.  Due to this brightness cutoff for spectroscopic observations, the aforementioned SN II-P with absolute \emph{r}-band magnitude of $-$16.0 would be included in our SCM sample only at redshifts of $z < 0.043$.  We also give priority to SNe that have good contrast with their host galaxy, which further biases against dimmer SNe.  When these observational limitations are considered, the relatively high intrinsic luminosity of our SN II-P sample - and thus the small intrinsic scatter - can be understood as the result of a magnitude limited search compounded by spectroscopic selection effects.

The small intrinsic scatter in \emph{I}-band magnitudes amongst our SNe II-P sample limits its utility for a self-contained determination of the velocity parameter $\alpha$.  The smaller the range of absolute magnitudes in a sample, the smaller the range of \fe\ velocities we expect to find.  This effect is compounded by the relatively large uncertainties in our day 50 \fe\ velocities; there are more SNe II-P with $\sigma_{v_{\mathrm{Fe\,II}}} > 400$~km~s$^{-1}$ in our sample of 15 SNe than in the entire P09 `culled' sample.  This partially explains the loose constraints that the SDSS data set is able to put on $\alpha$; both the results of P09 and no velocity dependence are consistent with our marginalized $\alpha$ at 2$\sigma$.  However, this effect is not able to explain why the intrinsic scatter increases when we apply the SCM to the combined sample of SDSS-plus-P09 SNe II-P.

To determine if we see any effects of bias as a function of redshift within our sample, we split our data into two roughly equal in size high- and low-redshift samples (redshift cut at $z = 0.065$) and determine the best-fit SCM parameters for each.  Although the resulting parameters have large uncertainties due to the low statistics, both sets of data have the same absolute magnitude and velocity parameters ($\alpha$ = 2.1 $\pm$ 0.8 and 1.9 $\pm$ 1.1 for low- and high-$z$ respectively). However, the value of the reddening term $R_I$ differs greatly; $R_I = 1.8 \pm 0.8$ for the low-$z$ sample, and $R_I = -0.4 \pm 0.4$ for the high-$z$ sample.  We reanalyze the low-$z$ SNe by systematically removing one object at a time and re-computing the $R_I$ parameter, and find that the high value of the low-$z$ reddening law parameter can be attributed to SN 2006iw, which has the smallest uncertainty in day 50 \VmI\ color of any object in our SCM sample (see Table \ref{SDSS-SCM}).  The $R_I$ parameter is effectively unchanged if any other low-$z$ SN is removed in the re-sampling analysis, but is consistent with zero when we remove SN 2006iw.  We find no evidence of bias in our fitted parameters as a function of redshift.  

We examine the effects of host galaxy extinction.  We expect to see fewer reddened objects at high redshifts; a given supernova appears brighter in a host galaxy with less extinction, and therefore is more likely to appear in our survey.  We do not find a strong trend of color as a function of redshift in our sample.  While our most distant SNe do have smaller \VmI\ values, the uncertainties are larger at these redshifts.  One aspect of our selection criteria for spectroscopic observation was that the SN be easily separated from its host galaxy; in effect, this requirement biases us against having heavily extincted, low-$z$ SNe II-P in our sample.  We also note that the colors of the SNe in our sample, though skewed towards the bluer end, fall within the range of observed values in P09.  The distribution of SN colors we observe and the resulting reddening parameter we obtain are both well explained by selection effects. 

It is reasonable to question whether selection effects that bias against fainter SNe being observed should even matter.  After all, the SCM is the \emph{Standardized} Candle Method; if a SN II-P sample consists of only bright SNe, these should all have high ejecta velocities, and therefore no bias should be introduced.  So, while the intrinsic magnitude dispersion without any corrections would appear tighter - and the average absolute magnitude brighter - than expected from a non-biased sample, the SCM should correct for this bias.  Malmquist bias alone is therefore unable to explain why we obtain a more luminous value of $M_{I_0}$ from the SDSS-SNS sample of SNe II-P than P09.  However, bias can still affect the analysis in a secondary manner if the measurement of the standardizing parameter is biased for the most luminous objects.  We examine whether the measurement of the ejecta velocity in our sample is biased towards lower values. 

As previously mentioned, the similarity in the derived $\alpha$ parameter at both low and high redshifts suggests that such a secondary bias, if one exists, is not a function of redshift.  Our spectra for high redshift SNe are primarily from larger telescopes, so the S/N of these spectra are not necessarily worse than those of low redshift SNe.  The distribution of uncertainties in the day 50 \fe\ velocities of the SDSS SNe II-P is flat as a function of redshift.  We do not appear to be biased against measuring ejecta velocity with less precision at high redshift.

\begin{figure}[htp]
  \centering
  \includegraphics[width=0.6\textwidth]{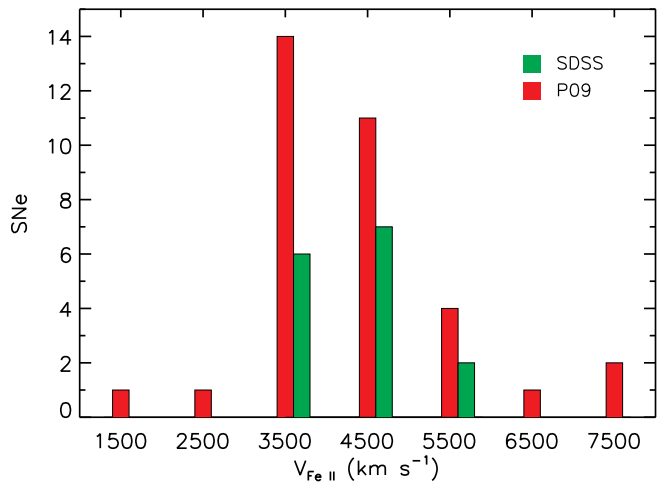}
  \caption{Distribution of ejecta velocity at day 50, measured through the cross-calibration of spectral templates to the \fe\ profile of an observed SN, for SNe II-P from the P09 `culled' sample (Red) and the SDSS-SNS (Green).  Each SN is placed in a bin of width 1000 km~s$^{-1}$, centered on the values shown.
\label{fig-vdist}}
\end{figure}

The distribution of day 50 ejecta velocities from our SN sample and the `culled' sample in P09, which were determined using the same method, is shown in Figure \ref{fig-vdist}.  The similarity of the distributions of these two data sets is difficult to reconcile with their different luminosity distributions within the context of the SCM.  One explanation for the difference could be that our day 50 magnitude derivation is systematically too luminous; since we have already shown that we expect our sample to be biased towards more luminous SNe, and the photometry for our sample is of excellent quality, this explanation seems unlikely (though we return to K-correction uncertainty later).   Another explanation of this effect could be that the measurement of our day 50 velocity is systematically offset to lower values.  Using the observed values of the \emph{I}-band magnitudes and \VmI\ colors from our sample of SN and assuming the SCM parameters from the `culled' sample of P09, we calculate the expected value of the day 50 \fe\ velocity for each SDSS SN II-P and compare it with our derived values.  The observed velocity is lower than that predicted by the P09 parameters by an average of $\sim 1300$ km~s$^{-1}$; only 3 of the 15 SNe are within $1\sigma$ of the predicted velocity.  We examine whether the low velocities of our bright sample could be caused by the N06 velocity extrapolation or the P09 cross-correlation method. 

Equation \ref{eq-vel} relates the velocity of the ejecta at any epoch from day 9-75 to that at day 50, in the middle of the plateau.  The distribution of epochs that we have spectra for is heavily skewed towards pre-day 50; only 1 SN in this study had a spectrum taken after day 50, and only a few had any taken post day 30.  If the relationship between the velocity at early times and day 50 is not accurately represented by Equation \ref{eq-vel}, we could be biasing our results relative to P09 and N06, which primarily have spectra from later epochs.  There are only 3 SNe in our sample for which we have one spectrum taken before day 20 and one at or after day 30, and in each one of these cases the extrapolated day 50 velocity from the earlier spectrum is lower than that derived from the later spectrum.  This would seem to imply that the epoch of the observation can bias the resulting day 50 \fe\ velocity.  However, there is no spectrum at any epoch in our dataset that has an \fe\ velocity high enough to match the prediction from the P09 set of parameters for the day 50 ejecta velocity of that SN; this suggests that our results cannot simply be ascribed to a bias in the N06 fitting function.  Furthermore, we applied Equation \ref{eq-vel} to each of the \fe\ velocities quoted in \citet{Quimby} for the very well observed Type II-P SN 2006bp, and found that for this object the direction of bias is reversed; the day 50 velocity obtained using the \fe\ velocities for the three spectra at $t < 20$ days is higher by $1000$ km~s$^{-1}$ than that obtained from each of the four velocity measurements at $t > 30$ days.  The accuracy of using Equation \ref{eq-vel} is therefore difficult to establish, as the velocity of the ejecta at early times may not have a one-to-one relationship with the velocity at day 50.

The cross-correlation method of P09 for deriving the \fe\ velocity from a SN II-P spectrum (D. Poznanski 2009, private communication) is applied in this paper to all of our SNe, as this should allow the most consistent comparison with the velocities derived for the full `culled' sample of P09.  However, as our spectra are typically from earlier epochs than in P09, it is possible that the cross-correlation is systematically underestimating the velocity at these times, leading to our low ejecta velocities.  We test for an early epoch bias by running the cross-correlation code on the SN 2006bp spectra obtained from the SUSPECT database, and compare the resulting velocities with those derived in \citet{Quimby} from the minima of the features in the Fourier transform-smoothed spectra.  There are two important aspects to our results.  First, the P09 method returns a much lower velocity for each epoch than is quoted in \citet{Quimby}, and the difference between the two methods is a strong function of epoch:  the P09 method differs by $300$-$400$ km~s$^{-1}$ at $t \geq 25$ days, while it is slower by at least $1500$ km~s$^{-1}$ at $t < 20$ days.  This difference between methods reinforces the notion that it is very difficult to measure accurately the velocity of a wide absorption feature, particularly at early epochs. However, since this method is used to measure velocities uniformly across datasets, the offset from the values of \citet{Quimby} should not be of importance to this work.  Second, using the velocities derived from the cross-correlation method, we find that the spectra at days 16 and 21 predict a higher day 50 \fe\ velocity for SN 2006bp than the day 42 and 57 spectra by $\sim 600$ km~s$^{-1}$; the early spectra overpredict the ejecta velocity in the plateau.  Another potential source of bias comes from the fact that early spectra are often dominated by H$\beta$ in the 4500-5500\AA\ range, so the cross-correlation could be matching this feature instead of \feII.  Since H$\beta$ is found at higher velocities than \feII, though, this effect could not result in our velocities being biased to lower values.  We thus find no indication that our velocities have been systematically offset to lower values due to the application of Equation \ref{eq-vel} or the P09 cross-correlation method.  Since introducing a different method of calculating the \fe\ velocities from our spectra is likely to provide biased results with respect to the data presented in P09 (as shown with SN 2006bp), we retain these two techniques as our spectroscopic analysis.  


The final source of uncertainty in our data is due to spectral templates.  As mentioned earlier, the SNe we observe are at high enough redshifts for K-corrections to be necessary, and we must perform S-corrections as well to place our data in the same filters as observed in previous studies.  These corrections require knowing the spectral energy distribution of each SN at the epoch (in our case, 50 days post explosion) we wish to correct - which we do not have.  Thus we must use a spectral template, which, though we warp the template to match observed colors, may not account for all variations between different SNe, and will be a source of error with unknown magnitude.  We are unable to create a template from our own data, as our spectroscopic follow-up program was biased towards early times, resulting in only 1 spectrum beyond day 40.  We use throughout this paper the same spectral template that was adopted for K-corrections in N06.  To test the extent to which warping the template can account for spectral differences between SNe, we obtained from the SUSPECT database a day 51 spectrum of SN 2004et and a day 42 spectrum of SN 2006bp.  Neither of these spectra cover the entire range of wavelengths necessary to perform our warp and K-corrections ($3500 - 11000$~\AA), so we scaled and pasted the N06 template to the blue and red end of each spectrum.  When we calculate the \emph{I}-band magnitudes for the SDSS SNe II-P using these new spectral templates, we find an average systematic uncertainty of 0.03 mag.  It is possible that with a larger database of spectra containing more diversity, we might find larger systematic uncertainties through the same test.  However, \emph{I}-band K-corrections using the spectrum of SN 2004et are 0.2 mag larger than the K-corrections using the N06 template and SN 2006bp when we do not warp the template.  The fact that the systematic uncertainty after warping is much smaller demonstrates that the act of warping the spectrum to the observed colors can correct for the majority of intrinsic dispersion amongst spectral templates.




\section{Discussion}
\label{sec-dis}

We have presented photometric and spectroscopic data for 15 SNe II-P from the SDSS-II Supernova Survey.  These SNe span a range of redshifts not covered by other surveys, connecting the local sample of HP02 and P09 to the high-$z$ SNLS sample in N06.  We propagate through our photometry reduction the uncertainty in the time of explosion in a way that accurately represents the state of knowledge regarding the explosion epoch.  We also perform K- and S-corrections for all of our SNe and describe the process in detail.  The best fit parameters for the SCM are updated with our new data, and are found to change significantly from those presented in P09.  The difference can be primarily attributed to the SDSS SNe being intrinsically brighter than the SNe in the `culled' sample of P09 (due to understood selection effects), without displaying the corresponding increase in ejecta velocity predicted by the SCM parameters of P09.  We show that the act of warping a SN II-P spectral template to the observed colors, while not completely removing systematic uncertainties, reduces them to such a level that they are unlikely to account for the magnitude of our offset.  The difficulty in accurately measuring the velocity of the broad \fe\ absorption features and the uncertainty involved in extrapolating velocities at early epochs to later times are likely to be the dominant sources of systematic uncertainty in our analysis.  The low $\alpha$ value we derive for the SDSS-only sample is due to both the small scatter in observed magnitudes and the uncertainty in our velocity determination.  Despite the selection effects and measurement uncertainties that limit the power of the SDSS SNe II-P to constrain the SCM parameters, we observe the primary relationship defining the SCM, which is that brighter SNe have faster ejecta, at $\approx 2\sigma$.  

The primary objective of the SDSS-II Supernova Survey was to identify and characterize SNe Ia; as a result, the observing strategy was not ideal for SNe II-P studies.  A future survey program that would focus on the SCM would greatly benefit from possessing the following qualities:

1)  Longer continuous survey duration.  For the final six weeks of such a survey, any newly discovered SNe will not have photometry sufficiently far into their plateau to allow extrapolation of their magnitudes at rest-frame day 50.  In addition, no SN can have its explosion date pinned down unless it happens after the first survey observation in its field.  Since each season of the SDSS-II SNS was three months long, no more than $50 \%$ of the survey time was useful for discovering new SNe II-P.  Doubling the length of such a survey would nearly triple the number of SNe discovered for which the SCM can be used.  

2)  Deeper, later spectroscopic observations.  When a new SN is discovered in a survey dedicated to SNe II-P, an initial quick spectrum is not needed.  A SN II-P will have a distinct light curve from a SN Ia, and thus valuable spectroscopic resources will not be used to follow up objects of no interest.  Since the velocity of the ejecta at day 50 is what is required, a spectrum at early times is not required, and observations can target day 50.  As we have shown, the systematic uncertainties associated with early-time spectra require a thorough study.  Also, each object must be observed for a sufficient time to obtain a high quality S/N spectrum.  Accurate measurements of \feII\ absorption lines are key to determining the strength of the $\alpha$ parameter; quality over quantity should be the mantra for these followups.

3)  Template database must be extended.  To perform both K-corrections and S-corrections, an accurate representation of the true spectral energy distribution is vital.  Ideally, the spectrum for each SN in such a survey would have sufficient wavelength coverage to allow corrections for all bands of interest to be computed from it directly.  As this situation is unlikely, complementary spectral observations that cover the near IR should allow for templates to be constructed that range from at least $4000 - 12000$\AA\, (or redder still, if the wavelengths in the lowest energy filter are redshifted out of this range at the targeted $z$).  The amount of uncertainty in the day 50 magnitude results that is due to relying on a limited number of SNe II-P spectral templates will only be known when more extensive templates are available. 

When these advances can be made, we will be able to determine whether the SCM and SNe II-P can place meaningful, independent constraints on cosmological parameters.

\acknowledgements
C.D. wishes to thank Dovi Poznanski and Peter Nugent for numerous helpful conversations.  We also wish to thank the anonymous referee for useful comments.  Funding for the SDSS and SDSS-II has been provided by the Alfred P. Sloan Foundation, the Participating Institutions, the National Science Foundation, the U.S. Department of Energy, the National Aeronautics and Space Administration, the Japanese Monbukagakusho, the Max Planck Society, and the Higher Education Funding Council for England. The SDSS Web Site is http://www.sdss.org/.

The SDSS is managed by the Astrophysical Research Consortium for the Participating Institutions. The Participating Institutions are the American Museum of Natural History, Astrophysical Institute Potsdam, University of Basel, University of Cambridge, Case Western Reserve University, University of Chicago, Drexel University, Fermilab, the Institute for Advanced Study, the Japan Participation Group, Johns Hopkins University, the Joint Institute for Nuclear Astrophysics, the Kavli Institute for Particle Astrophysics and Cosmology, the Korean Scientist Group, the Chinese Academy of Sciences (LAMOST), Los Alamos National Laboratory, the Max-Planck-Institute for Astronomy (MPIA), the Max-Planck-Institute for Astrophysics (MPA), New Mexico State University, Ohio State University, University of Pittsburgh, University of Portsmouth, Princeton University, the United States Naval Observatory, and the University of Washington.

The Hobby-Eberly Telescope (HET) is a joint project of the University of Texas at Austin, the Pennsylvania State University, Stanford University, Ludwig-Maximillians-Universit\"at M\"unchen, and Georg-August-Universit\"at G\"ottingen.  The HET is named in honor of its principal benefactors, William P. Hobby and Robert E. Eberly.  The Marcario Low-Resolution Spectrograph is named for Mike Marcario of High Lonesome Optics, who fabricated several optics for the instrument but died before its completion; it is a joint project of the Hobby-Eberly Telescope partnership and the Instituto de Astronom\'{\i}a de la Universidad Nacional Aut\'onoma de M\'exico.  The Apache Point Observatory 3.5-meter telescope is owned and operated by the Astrophysical Research Consortium.  We thank the observatory director, Suzanne Hawley, and site manager, Bruce Gillespie, for their support of this project.  The Subaru Telescope is operated by the National Astronomical Observatory of Japan;  we thank Takashi Hattori for his support during our observing run at Subaru.  The William Herschel Telescope is operated by the Isaac Newton Group, the Nordic Optical Telescope is operated jointly by Denmark, Finland, Iceland, Norway, and Sweden, and the Telescopio Nazionale Galileo (TNG) is operated by the Fundaci\'on Galileo Galilei of the Italian Istituto Nazionale di Astrofisica (INAF), all on the island of La Palma in the Spanish Observatorio del Roque de los Muchachos of the Instituto de Astrof\'{\i}sica de Canarias.  Observations at the ESO New Technology Telescope at La Silla Observatory were made under programme {\small ID}s 77.A-0437, 78.A-0325, and 79.A-0715.  Kitt Peak National Observatory, National Optical Astronomy Observatory, is operated by the Association of Universities for Research in Astronomy, Inc. (AURA) under cooperative agreement with the National Science Foundation.  We thank R. Kirshner, P. Challis, and S. Blondin for assistance with the spectroscopic observations at Magellan.  The Hiltner 2.4-m telescope of the MDM Observatory is owned and operated by the University of Michigan, Dartmouth College, the Ohio State University, Columbia University, and Ohio University.

\clearpage

\end{document}